\documentclass[fleqn,usenatbib]{mnras}
\usepackage{graphicx}
\usepackage{epstopdf}

\usepackage{newtxtext,newtxmath}
\usepackage[T1]{fontenc}
\DeclareRobustCommand{\VAN}[3]{#2}
\let\VANthebibliography\thebibliography
\def\thebibliography{\DeclareRobustCommand{\VAN}[3]{##3}\VANthebibliography}

\title[Spin evolution after DWD merger]{Spin evolution modeling  for  a newly-formed white dwarf resulting from  binary  white dwarf
  merger}

\author[Cheng Y. \& Takata J.]{
Yanchang  Cheng,$^{1}$\thanks{E-mail: chengyc@hust.edu.cn}
and Jumpei Takata,$^{1}$\thanks{E-mail: takata@hust.edu.cn}
\\
$^{1}$Department of Astronomy, School of Physics, Huazhong University of Science and Technology, Wuhan, Hubei 430074, China\\
}

\date{Accepted XXX. Received YYY; in original form ZZZ}

\pubyear{---}
\begin{document}
\label{firstpage}
\pagerange{\pageref{firstpage}--\pageref{lastpage}}
\maketitle

\begin{abstract}
  Merger of two white dwarfs (WDs)  has been proposed to form an isolated WD having high magnetization and rapid rotation. We study the influence of the  magnetohydrodynamic (MHD) wind on spin evolution of the newly-formed merger
  product. We consider  the scenario  that the merger product  appears as a  giant-star-like object
  with a radius of $>10^{10}~{\rm cm}$ and a luminosity of the order of an Eddington value.
  We  solve a structure of the merger product under the  hydrostatic equilibrium and identify the position of the slow-point in the hot envelope.  It is found
  that  if such a giant-star-like object is spinning with an angular speed of the order of the Keplerian value, the MHD wind can be produced.  The mass-loss rate is estimated to be of the order of $\sim 10^{20-21}~{\rm g~s^{-1}}$, and the timescale of the spin down is
  $\sim 10-10^{3}$~years, which depends on stellar magnetic field.  We discuss  that  the final angular momentum when
  the MHD wind is terminated is related to the magnetic flux and initial radiation luminosity of the merger product.
  We apply our model to  three  specific magnetic  WD sources ZTF J190132.9+145808.7, SDSS J221141.8+113604.4, and PG 1031+234 by assuming that those WDs were as a result of the merger product.  We argue that
  the current periods of ZTF J190132.9+145808.7 and PG 1031+234 that are strongly magnetized WDs are related to the initial luminosity at the giant phase. For
  SDSS J221141.8+113604.4, which is mildly magnetized WD, its angular momentum  was almost determined when the spin-down timescale due to MHD wind  is comparable to the cooling timescale in the giant phase.

\end{abstract}


\section{Introduction} \label{sec:intro}
White dwarf (hereafter WD) is the end product of a main-sequence star (hereafter MS) with a mass less than $\sim 8~M_{\odot}$ and
it is the most common endpoint of the stellar evolution; for example,~\cite{korol2020populations} estimate that
total number of the WD in the Milky Way is estimated to be $\sim 3\times 10^9$.
A WD can be produced not only as a result of the single stellar evolution but also as a result
of a merger of two stars in the binary system~\citep{badenes2012merger,shen2012long,temmink2020looks}.
It has been estimated that about 10-30\% of the single WDs in the Milky Way are the products of the binary merger~\citep{badenes2012merger,maoz2017binary,cheng2020double,temmink2020looks}.
Recent {\it Gaia} observation is drastically increasing the population of known WDs~\citep[and reference therein]{tremblay2024gaia},
and has also collected evidences of the WD produced by the binary merger process~\citep{cheng2019cooling,cheng2020double,kilic2023merger}.
For example,~\cite{cheng2020double} find that the distribution of the transverse velocity of the massive WD with $(0.8-1.3)~M_{\odot}$  deviates significantly 
from the case predicted by the WD produced via the single stellar evolution and they suggest that the additional production process of the massive WD is required.~\cite{temmink2020looks} use the binary evolution synthesis code for the production of the WD through the various type binary merger (MS+MS binary, MS+WD binary, WD+WD binary, etc), and demonstrate that the WD+WD binary merger (hereafter DWD merger) is the dominant process to create the massive WDs ($>0.9~M_{\odot}$).
It has also been proposed that the DWD merger produces a highly magnetized WD~\citep{wickramasinghe2000magnetism,garcia2012double,ji2013post,hernandez2024rotation}. In particular,
the recent observations have revealed a number of {\it isolated} massive, fast-spinning and high-field  WDs that were probably produced through the WD+WD binary channel.
SDSS J221141.8+113604.4 (hereafter J2211+1136)  is one of the fastest spinning isolated WD with a spin period of $P_{WD}\sim 70.32$~s,
a stellar mass of $M_{WD}\sim 1.27~M_\odot$ and a dipole field strength of $B_{WD}\sim 1.5\times 10^{7}$~G~\citep{kilic2021isolated}.  ZTF J190132.9+145808.7 (hereafter J1901+1458) has a spin period of $P_{WD}\sim 416.4$~s, a mass of $M_{WD}\sim (1.327-1.365)~M_{\odot}$ and a magnetic field of $B_{WD}\sim  (6-9)\times 10^8$~G~\citep{caiazzo2021highly}.
WD J005311 is the WD identified in a mid-infrared nebula that is thought to be a result of a  DWD merger event~\citep{gvaramadze2019massive}. This is
because its effective temperature (around $2\times10^5$~K) and the feature of deficiency of hydrogen and helium from the central star and nebula agree with the evolution of
a DWD merger product predicted by~\cite{schwab2016evolution}. \cite{kashiyama2019optically} propose that the mid-infrared nebula around the WD J005311 is powered by a fast-spinning and high-field WD ($P_{WD}\sim 2-5$~s and $B_{WD}\sim (2-5)\times 10^7$~G). Such a fast-spinning and highly magnetized  WD may accelerate the charged particles in the magnetosphere and produce non-thermal radiation, as the X-ray emission from J1901+1458 indicates~\citep{bamba2024x}.  These rapidly spinning magnetized WD also gives a hint of the origin of  classical  magnetic WD with 
a longer spin period and a smaller mass, such as 
PG 1031+234 ($M_{WD}\sim0.96~M_{\odot}$,  $P_{WD}\sim212$~min  and  $B_{WD}\sim (2-10)\times 10^8$~G)~\citep{schmidt1986new,brinkworth2007survey,gentile2021catalogue}.  If these magnetic WDs are product of the DWD merger, a unified picture of the spin-evolution after the merger will be necessary to understand the difference in the spin periods among the magnetic  WDs.

In this study, we assume that the DWD merger events produce the isolated magnetized WDs. It has been suggested that the merger product  appears with  a luminosity of the order of  Eddington value, $L_E\sim 3.2\times 10^4 L_{\odot} (M_{WD}/M_{\odot})$~\citep{wu2022formation}. The cooling age of the WD radiating luminosity of $L_{WD}$ may be estimated from \citep{koester1990physics},
\begin{equation}
  \tau_{cool}\sim \frac{10^8}{A}\left(\frac{M_{WD}}{M_{\odot}}\right)^{0.609}\left(\frac{L_{WD}}{L_{\odot}}\right)^{-0.609}~{\rm years},
  \end{equation}
where $A$ is the atomic mass number. Assuming CO core ($A\sim 12$), the current cooling age is estimated to be $\tau_{cool}\sim 3$~Gyr for J2211+1136~\citep{kilic2021most}, $\tau_{cool}\sim 0.075$~Gyr for J1901+1458 with the observed luminosity of $L\sim 0.03~L_{\odot}$~\citep{caiazzo2021highly}, while $\tau_{cool}\sim 0.1$~Gyr for PG 1031+234 with $L\sim 0.01~L_{\odot}$~\citep{gentile2021catalogue}. On the other hand, by  assuming  that the measured magnetic field, $B_{WD}$, represents the dipole magnetic field of the WD,  we may estimate the spin-down timescale
due to the dipole radiation as
\begin{eqnarray}
  \tau_{sd}&=&\frac{3}{(2\pi)^2}\frac{c^3P_{WD}^2I_{WD}}{R_{WD}^6B_{s}^2\sin^2\chi} \nonumber \\
  &\sim& 0.5~{\rm Gyr}\left(\frac{P_{WD}}{100~{\rm s}}\right)^2
  \left(\frac{B_{WD}}{10^8~{\rm G}}\right)^{-2} \nonumber \\
&&  \left(\frac{R_{WD}}{5\cdot10^8~{\rm cm}}\right)^{-6}
  \left(\frac{I_{WD}}{10^{50}~{\rm g~cm^2}}\right)\sin^{-2}\chi,
  \label{eq:dipole}
\end{eqnarray}
where $\chi$ is the inclination angle of the magnetic axis, $R_{WD}$ is the radius of the WD, and $I_{WD}$ is the moment of inertia. We can see that with the estimated mass  $M_{WD}\sim 1.3~M_{\odot}$ and radius $R_{WD}\sim 3\times 10^{8}$~cm,
the spin-down timescale of J2211+1136 ($P_{WD}\sim 70.32$~s and $B_{WD}\sim 1.5\times 10^{7}$~G)
becomes  $\tau_{sd}\sim 200$~Gyr, which is longer than the cooling timescale. 
We can also see that the spin-down timescales due to the dipole radiation of J1901+1458 and ($>10$~Gyr) and of PG 1031+234  ($>100$~Gyr) are also  longer than  their cooling timescales. Although the true ages of the WDs are unknown, the effect of the dipole radiation is not primary mechanism for achieving the observed spin periods.  This suggests that (i) those three  WDs were produced with the current spin angular momentum, or (ii) a spin-down torque by other processes has operated.

The spin evolution of the  merger product  has been discussed by previous studies. 
 In the models discussed in  \cite{kulebi2013magnetic}   and \cite{sousa2022double}, for example,  the main cause of the spin-down of the merger product is an interaction of the magnetic field of the product with a debris disk, which is remnant of the disrupted WD~\citep{shen2012long,ji2013post}. \cite{kulebi2013magnetic}   apply their model to the three magnetic WDs (WDs 0325-857, 1015+014 and PG 1031+234)  with a spin period of $\sim 10-200$~min and they suggest that the current spin periods are close to the values when the accretion was switched off in a timescale of $\sim 10^{4-5}$~years. On the other hand,
   \cite{shen2012long} study the long-term evolution of the remnant of the DWD merger
   and suggest that the magnetic stress redistributes the angular momentum from the core to the Keplerian disk and quickly establishes a rigidly rotating  and giant-star-like object, which is not surrounded by the accretion disk~\citep{2012MNRAS.427..190S,schwab2021evolutionary}. \cite{schwab2021evolutionary} estimates a loss of the angular momentum during the giant phase, which is supposed to last of the order of $10^4$~years after the merger event, and predicts that the new born WD has an initial period of $10-20$~minutes.
     
   In this paper, we revisit the effect of the stellar wind on the spin evolution of the hot product born after the DWD merger, since the previous studies have not investigated this effect well. In particular, we will determine the position of the slow-point and the mass-loss rate of the stellar wind by solving the structure of the
   merger product under the assumption of the  hydrostatic equilibrium. 
   In section~\ref{sec:2}, we introduce our model for the magnetized wind and structure of the merger product. We discuss the condition of 
   the launch of the wind and the angular momentum loss by the wind. In section~\ref{sec:3}, we discuss the general picture of the spin-down of the merger product by the magnetized wind  and    apply our model to the three  WDs, J1901+1458, J2211+1136
   and PG 1031+234, as representative examples.  We compare our model with the previous studies   in section~\ref{sec:4}
   and summarize our results in section~\ref{sec:5}.

   We note that  the merger product may not be considered as a conventional 
   WD in its initial evolution stage, because
   the envelope with  a mass of $M_{env}\sim 0.1-0.3M_{\odot}$ is sustained by the gas and/or radiation pressures, and it may have a radius of  $>10^{10}$~cm. However, we use the subscript ``WD'' to represent the physical quantities of the merger product, since the remnant eventually evolves to a WD after cooling down sufficiently.

\section{Theoretical model} 
\label{sec:2}
\begin{figure*}
  \includegraphics{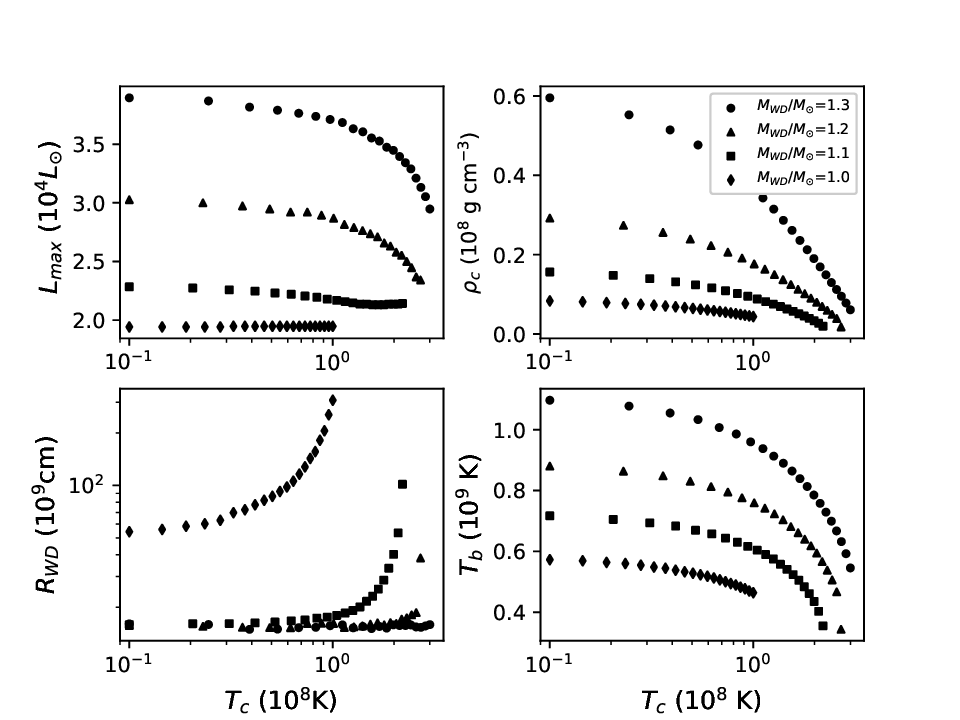}
  \caption{Properties of the merger product with $L_{WD}\sim L_{\rm max}$ as a function of the core temperature.
    Top-left: The luminosity above which the energy transfer
    via the convection may be important in the envelope. Our current calculation cannot be applicable to
    $L_{WD}\ge L_{\rm max}$. Top-right: Central density. Bottom-left: Radius of the merger product.
    Bottom-right: Temperature at the core/envelop boundary.  The circle, triangle, square and diamond symbols show
    the cases for the mass $M_{WD}/M_{\odot}=1.3$,
    1.2, 1.1 and 1.0, respectively. The results are envelope mass of $M_{env}=0.3M_{\odot}$ for all cases. }
  \label{fig:in}
\end{figure*}

\begin{figure*}
  \includegraphics[width=0.9\linewidth]{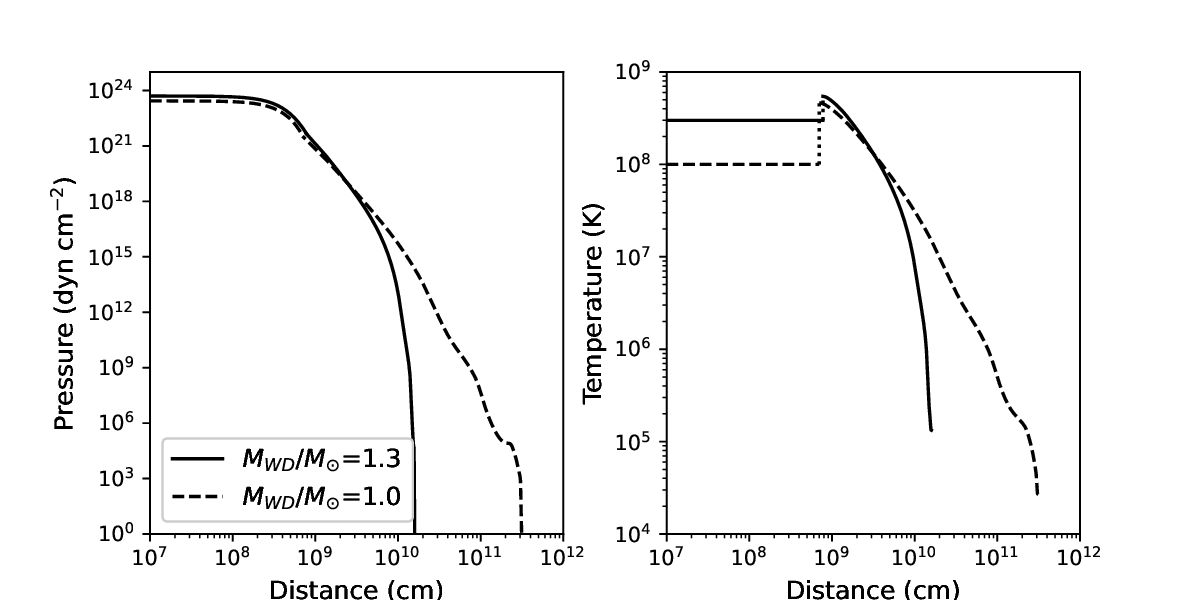}
  \caption{Profiles of the pressure (left panel) and temperature (right panel) for the merger products with the
    mass $M_{WD}/M_{\odot}=1.3$ (solid line) and 1.0 (dashed line), respectively. The
    luminosity and core temperature  are $L_{WD}=3\times 10^4L_{\odot}$ and $T_c=3\times 10^8$K, respectively,
    for $M_{WD}/M_{\odot}=1.3$
    and $1.9\times 10^4L_{\odot}$ and $T_c=10^8$K, respectively, for  $M_{WD}/M_{\odot}=1.0$. The envelope mass
    is $M_{env}=0.3M_{\odot}$.}
  \label{fig:structure}
\end{figure*}

\subsection{Magnetohydrodynamic wind and angular momentum loss}
\label{sec:2.1}
We assume that a DWD merger creates a rapidly spinning and highly magnetized object. It has been predicted that initial stage of the merger product  is very hot and produces a radiation luminosity of the order of the Eddington value~\citep{shen2012long,schwab2016evolution,wu2022formation}. Such a hot merger product may launch a magneto-hydrodynamics (MHD) wind that carries away the angular momentum of the magnetized object. An exact treatment of the MHD  wind is a complicated physical process and requires a treatment of the numerical simulation~\citep{kashiyama2019optically,zhong2024optically}. In our study, therefore,  we simplify the treatment of the stellar wind to investigate a basic picture of the spin evolution of the merger product if the MHD wind would be launched.  We explore a steady wind, since the spin-down timescale under the effect of the wind is in general longer than
the dynamical timescale of the wind, which may be defined by $R_A/V_A$ with $R_A$ being the radius of the Alf\'{v}en point and $V_A$ the wind  speed at the Alf\'{v}en point. We assume that the magnetic field is a mono-pole type structure in the poloidal plane, and discuss the wind solution in the equatorial plane.  Such a wind solution is greatly discussed in previous studies of the general stellar wind~\citep{weber1967angular, michel1969relativistic,lamers1999introduction,mestel2012stellar}. In the non-relativistic MHD wind,
the equation of motion in the radial direction ($r$) may  be described as
\begin{equation}
  V_{r} \frac{dV_{r}}{dr}+\frac{1}{\rho}\frac{dP_i}{dr}+\frac{Gm}{r^2}-\frac{\kappa L}{4\pi r^2 c}-\frac{V_\phi^2}{r}+\frac{B_\phi}{4\pi
    \rho r}\frac{d}{dr}(rB_\phi)=0,
    \label{eq:dynamic}
\end{equation}
where $m$ is the stellar mass within the radius $r$ and $L$ is the radiation luminosity at the radial distance $r$. In addition,  $V_r$ and $V_{\phi}$ are the radial and azimuth speeds of the wind respectively, $P_i$ is the gas pressure, $\rho$ is the mass density of the wind, $\kappa$ is the opacity,  and $B_\phi$ is the azimuthal component of the magnetic field. The fourth term on the left hand side corresponds to the radiation pressure.
A steady MHD wind passes through the slow-, Alf\'{v}en- and fast-points, and
it can be launched when  the envelope of the merger product satisfies the
condition of the slow-point. To identify the slow-point, we may rewrite the equation~(\ref{eq:dynamic}) in a form of~\citep{kashiyama2019optically},
\begin{eqnarray}
\left(V_r^2-c_s^2-\frac{A_\phi^2V_r^2}{V_r^2-A_r^2}\right)&\times& \frac{r}{V_r}\frac{dV_r}{dr}=\frac{\kappa L}{4\pi rc}
+\frac{k_B}{\mu m_p}\left(-r\frac{dT}{dr}+2T\right)
  \nonumber \\ 
&&  -\frac{Gm}{r}+V_\phi^2+2V_rV_\phi\frac{A_rA_\phi}{V_r^2-A_r^2},
  \label{eq:dynamic1}
\end{eqnarray}
where $T$ is the temperature,  $c_s=\sqrt{k_BT/(\mu m_p)}$
 is the sound speed of the ideal gas  with $k_B$ being the Boltzmann constant,  $\mu$ the mean atomic weight and $m_p$  the proton mass. In addition,  $A_{r}$ and $A_{\phi}$ are local Alf\'{v}en speeds corresponding to
the radial  and azimuthal magnetic fields, respectively. We anticipate that at the slow-point, the  Alf\'{v}en speed is much  slower
than the sound speed, $A_{r, \phi}<< c_s$, and the rotation velocity of the  wind is nearly co-rotation speed
($V_\phi=2\pi r/P_{WD}$). Applying these approximations on the equation~(\ref{eq:dynamic1}),
the condition of the slow-point may be described as ~\citep{lamers1999introduction},
\begin{equation}
 F(r)\equiv \frac{\kappa L}{4\pi rc}
 +\frac{k_B}{\mu m_p}\left(-r\frac{dT}{dr}+2T\right)-\frac{Gm}{r}
   +r^2\Omega_{WD}^2,
  \label{eq:func}
\end{equation}
and
\begin{equation}
  F(R_s)=0~{\rm and}~V_r(R_s)=c_s(R_s)
   \label{eq:slow}
  \end{equation}
where $R_s$ represents the radial distance to the slow-point and  $\Omega_{WD}=2\pi/P_{WD}$ is the angular speed of the WD's spin. We evaluate the mass-loss rate of the wind  at the slow-point as
\begin{equation}
  \dot{M}_W=4\pi R_s^2\rho(R_s)c_s(R_s).
  \label{eq:massloss0}
  \end{equation}

The angular momentum carried by the MHD wind is determined at the Alf\'{v}en-point.  The rate
of the angular momentum loss per unit time due to the wind may be expressed by \citep{mestel1999stellar}
\begin{equation}
  \dot{J}_{w}=\frac{3\dot{M}_Wc^2}{2\Omega_{WD}}\left(\frac{V_M}{V_f}+\frac{V_M^3}{c^2V_f}-1\right),
  \label{J_dot}
\end{equation}
where $V_M=(\Phi_B^2\Omega_{WD}^2/\dot{M}_W)^{1/3}$ is the so-called Michel speed, $\Phi_B=r^2B_r={\rm constant}$ is the radial magnetic flux,
 and   $V_f=c\sigma^{1/3}(1+\sigma^{2/3})^{-1/2}$ with $\sigma=V_M^3/c^3$ is the terminal wind speed. In the limit of $\sigma<<1$,
the equation~(\ref{J_dot}) becomes $\dot{J}_{w}\sim \dot{M}_W V_M^2/\Omega_{WD}$ of the non-relativistic case.  The distance to  the Alf\'{v}en-point may be approximately represented
as  $R_A= (c/\Omega_{WD})\{(3\Omega_{WD}\dot{J}_w/2\dot{M}_Wc^2)/[1+(3\Omega_{WD}\dot{J}_w/2\dot{M}_Wc^2)]\}^{1/2}$, which is $R_A\sim\sqrt{3/2} V_M/\Omega_{WD}$ in the
non-relativistic wind regime and $R_A=c/\Omega_{WD}$ in the relativistic wind regime.
The Alf\'{v}en speed at the Alf\'{v}en radius is expressed as $(V_A/c)=\gamma_c/\sqrt{1+\gamma_c^2}$
with $\gamma_c=\gamma_AV_A/c=[1-(\Omega_cR_A/c)^2](c/\Omega_{WD} R_A)^2\sigma$, which becomes
$V_A\sim(2V_M/3)$ in the non-relativistic wind  and $V_A\sim 0.55~c$  in the relativistic wind.

For the case of the weakly magnetized merger product, the  Alf\'{v}en-point, to which  the radial distance  is described by 
$R_{A}\sim \sqrt{3/2}V_M/\Omega_{WD}\propto \Phi_B^{2/3}$,  may be located inside  the slow-point, and the angular momentum loss by equation~(\ref{J_dot}) will not be applicable. In such a situation, we anticipate that the
torque from the stellar  wind exerts on  the stellar surface  and
we calculate the angular momentum loss from
\[
\dot{J}_{W}=\frac{3}{2}\dot{M}_WR_{WD}^2\Omega_{WD}.
\]

The treatment of the MHD wind becomes inapplicable once the mass-loss rate falls to  value, with which the number density at
the Alf\'{v}en-point becomes of the order of the Goldreich-Julian number density of
$n_{GJ}(R_A)\sim B(R_A)\Omega_{WD}/2\pi ce$ \citep{goldreich1969pulsar}.
Such a lower density magnetosphere will induce a large electric field due to the rotation of
the magnetized star, and may finally establish a charge-separated force-free structure. The mass-loss rate at the boundary between
the two states of the magnetosphere will be of the order of $\dot{M}_W\sim 4\times 10^{10}(\Phi_B/10^{26}~{\rm G~cm^2})(P_{WD}/100~{\rm s})^{-1}~{\rm g~s^{-1}}$. As discussed in section~\ref{sec:estimation}, on the other hand, we will find that the typical  mass-loss rate from the merger product is of the order of $\dot{M}_W\sim 10^{20-21}~{\rm g~s^{-1}}$, suggesting
that the treatment of the MHD wind is applicable for the current study. 

\subsection{Structure of merger product}
\begin{figure}
  \includegraphics[width=1\linewidth]{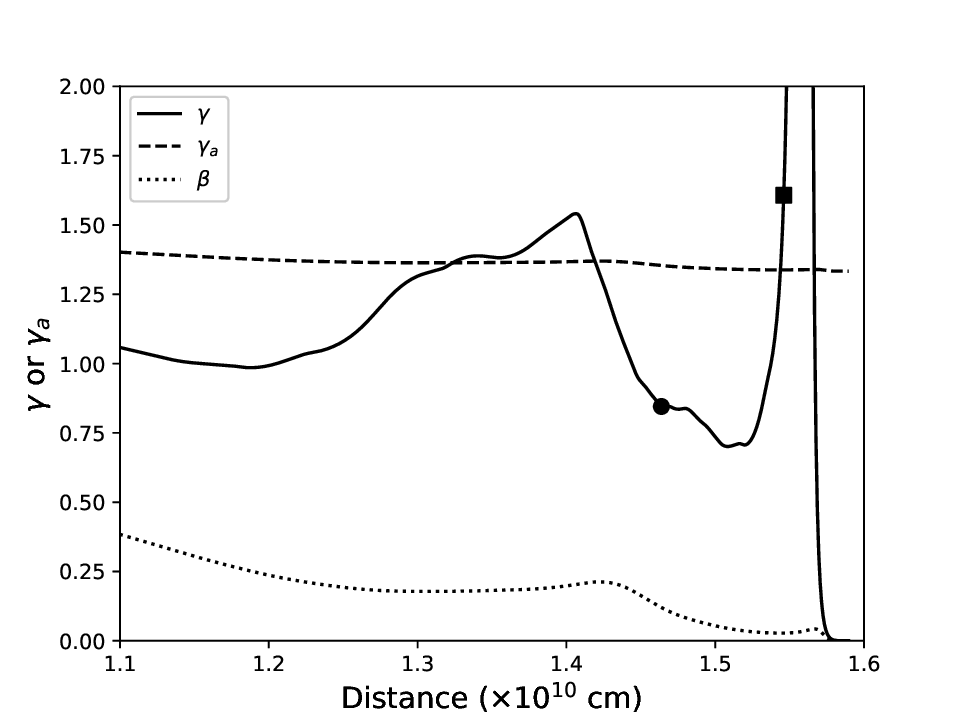}
  \caption{ Examining  Schwarzschild criterion of the convection in the envelope.
      The results are the structure of $M_{WD}=1.3M_\odot$ in Figures~\ref{fig:structure}.  
     The solid line is the index defined by $\gamma\equiv (\rho/P)|dP/d\rho|_{star}$ and the dashed line is the adiabatic index calculated from equation~(\ref{ga}).  The dotted line represents $\beta=P_i/P$. The filled circle and square represent the position of the slow-point and the photosphere, respectively. }
  \label{fig:ga}
\end{figure}
\begin{figure*}
  \includegraphics[scale=0.8]{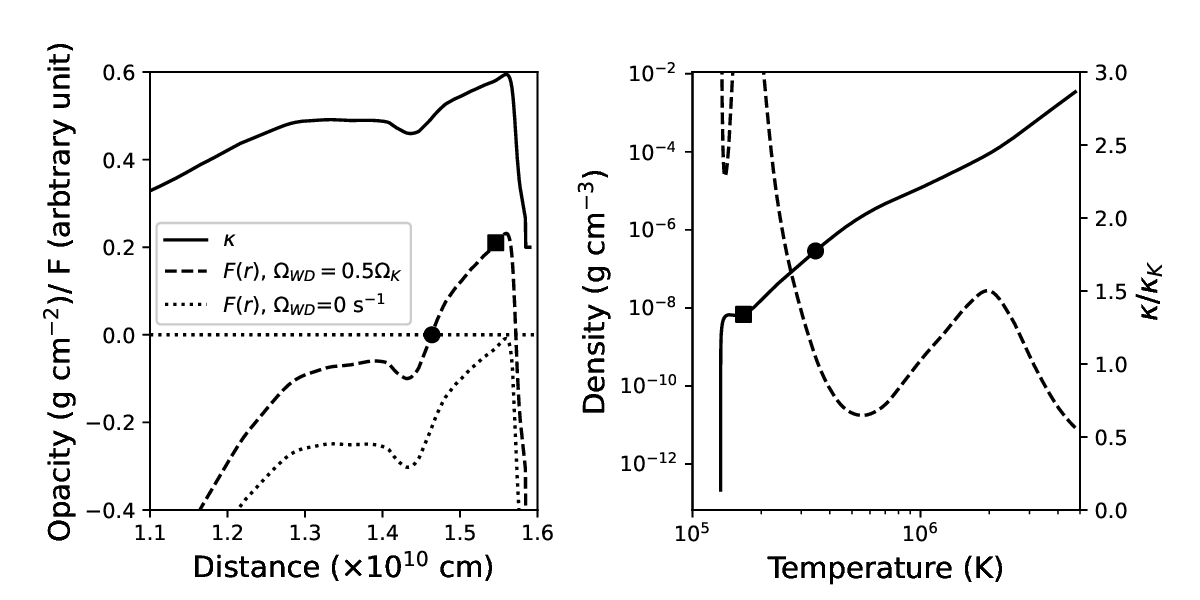}
  \caption{ Left: Profile of the opacity (solid line) and  the function $F$ of equation~(\ref{eq:func}) with 
  the angular frequency of $\Omega_{WD}=0.5\Omega_{K}$ (dashed line) or  no rotation (dotted line). The point satisfying both $F(r)=0$ and $dF(r)/dr>0$ is  identified as the slow-point (filled circle). The filled square presents 
    the position of the photosphere. 
    The result is the case for the mass $M_{WD}=1.3M_{\odot}$ with the parameters same as those in Figure~\ref{fig:structure}.  Right: Relation between  mass density and temperature (solid line) within radial
    distance corresponding to the left panel. The dashed line is $\kappa/\kappa_{K}$, where
  $\kappa$ is the opacity defined by equation~(\ref{eq:opacity}) and $\kappa_{K}$ is the Kramer formula defined by equation~(\ref{eq:kramer}). }
  \label{fig:opa}
  \end{figure*}

\begin{figure}
\includegraphics[width=1\linewidth]{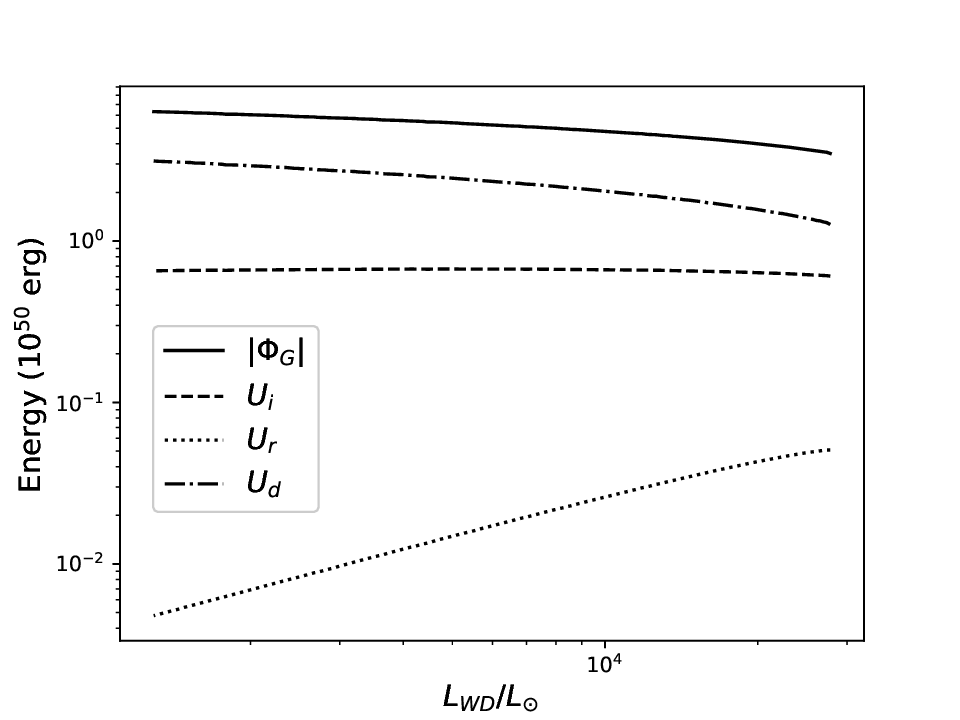}
  \caption{ Energy of the merger product as a function of the radiation luminosity. The solid line is the absolute value of the gravitational potential  energy, $|\Phi_G|$. The dashed, dotted and dashed-dotted lines represent the internal
      energies of the ideal gas ($U_i$), the radiation ($U_{r}$) and degenerate gas ($U_d$), respectively.
      The results are case for  the stellar mass of
      $M_{WD}/M_{\odot}=1.3$ and the envelope mass of $M_{env}=0.3M_{\odot}$.}
  \label{fig:grav}
\end{figure}

\begin{figure*}
  \includegraphics[width=0.9\linewidth]{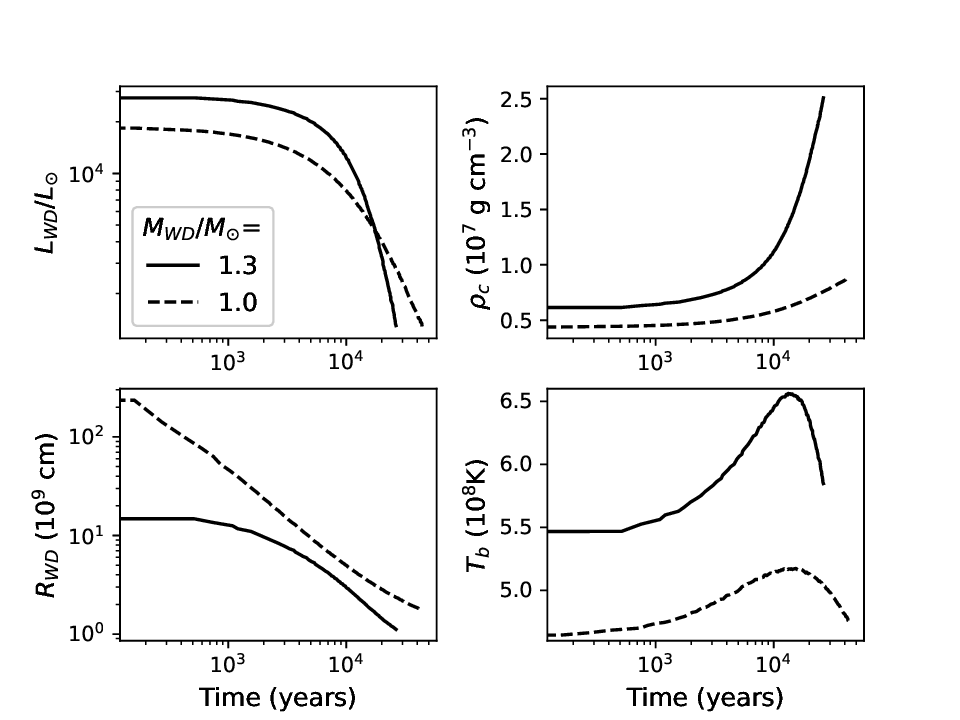}
  \caption{ The properties of the merger product as a function of the cooling age. Top-left: Luminosity. Top-right: Central mass density. Bottom-left: Radius of the merger product. Bottom-right: Temperature at the core/envelope boundary. The solid and dashed lines are the results for  $M_{WD}=1.3M_{\odot}$ with $T_c=3\times 10^8$~K and $1.0M_{\odot}$
      with $10^8$~K, respectively. The mass of the envelope is $M_{env}=0.3M_{\odot}$. }
  \label{fig:cool}
\end{figure*}

\subsubsection{Assumption and approximation}
\label{sec:assumption}
Previous studies have suggested that the product after the merger is  composed of a cooler core surrounded  by a hotter envelope~\citep{dan2014structure, schwab2016evolution, schwab2021evolutionary, wu2022formation}. The cooler core is remnant of the primary WD before the  merger, and  the hot envelope is mainly originated from
the disrupted WD.  The hot envelop can  produce a radiation  luminosity of the order of the Eddington value and can expand  to a radius of $r>10^{10}$~cm, which is so-called
giant phase of the merger product. In this study, we will evaluate the spin-down  as a result of the MHD wind from the giant-star-like
 merger product.

To launch the MHD wind, the solution of the MHD equations must  satisfy the condition~(\ref{eq:slow})
in the envelope of the merger product. It is discussed that  the structure of the subsonic region ($r<R_{s}$) is mainly  determined by the hydrostatic structure, because the dynamic pressure is smaller than gas/radiation pressure~(see figure~3.2 in \cite{lamers1999introduction}). In this study, therefore, we identify the slow-point by solving the hydrostatic structure of the non-rotating star. The effect of the stellar spin on the structure of the envelop will become an  important factor  when the stellar angular speed is of the order of the Keplerian value,
\begin{equation}
\Omega_K\equiv\left(\frac{GM_{WD}}{R^3_{WD}}\right)^{1/2}.
\end{equation}
Nevertheless,  we do not taken into account the effect of the spin for simplicity. Since the merger product  is quickly  spun down  once the wind is launched, the final spin period, at which the  condition of the slow-point is unsatisfied  in the envelop,  can be significantly  smaller than the Keplerian value.

As presented in section~\ref{hstru}, we will solve the hydrostatic structure of the star for the specific radiation luminosity ($L_{WD}$) under the treatment of the {\it radiative} energy transfer. As the assumed luminosity increases, the outer radius of the
envelop also increases.  On the other hand, if the heat flux is very large, the calculated gas pressure ($P_i$)
at some region of the envelope increases with the radial distance, namely  $dP_i/dr>0$. In such a case, the assumption of the radiative energy transfer will be violated, and the convection process will describe the heat transfer. 
Since treatment of the convection process is complex,  we restrict the radiation luminosity of $L_{WD}<L_{\rm max}$, where $L_{\rm max}$ is the critical luminosity above which the treatment of the  radiative energy transfer is violated. The critical luminosity
is of the order of the Eddington luminosity of 
\begin{equation}
L_{\rm max}\sim \frac{4\pi GM_{WD}c}{\kappa_{\rm max}},
\end{equation}
where $\kappa_{\rm max}$ is the maximum value of the opacity in the envelope.  In section~\ref{sec:3.1}, we will evaluate  the stability against convection using the Schwarzschild criterion.

The structure of the envelope depends on  the main compositions through the opacity, $\kappa$. In this study, we will consider the main composition of the merger product is a mixture of the Carbon (C) and Oxygen (O), and  calculate the opacity from 
\begin{equation}
  \kappa=\frac{\kappa_{\rm con}\kappa_{\rm rad}}{\kappa_{\rm con}+\kappa_{\rm rad}},
\label{eq:opacity}
\end{equation}
where $\kappa_{\rm con}$ and $\kappa_{\rm rad}$ correspond to  the conductive  and radiative  opacities, respectively. We use the conductive
opacity\footnote{\url{http://www.ioffe.ru/astro/conduct/}} presented in
\cite{2007ApJ...661.1094C} and \cite{2015SSRv..191..239P}. 
For the radiative opacity, we use the open code\footnote{\url{https://www.cita.utoronto.ca/~boothroy/kappa.html}}  for  OPAL stellar opacity~\citep{1996ApJ...464..943I}. We use the OPAL table for the enhanced C and O elements and apply $X_{\rm C}\sim 0.5$ and $X_{\rm O}\sim 0.5$, respectively.

It has been suggested that the core of the primary is compressed during the merger process and its temperature can  reach to $T_c\sim (1-3)\times 10^8$~K~\citep{2016MNRAS.462.2486F}. However,  since it has not been well investigated how the core temperature after the  merger  depends on the composition and mass, the previous studies assume  the core temperature of  $T_c=10^{7-8}$~K. In this study, we also treat the core temperature as a model parameter.

Inside the merger product, the temperature reaches  the peak value at the region around the core/envelope boundary, as demonstrated in, for example,
\cite{schwab2021evolutionary} and \cite{wu2022formation}. Such a temperature profile
will cause a heat transfer from the hotter envelope to the cooler core.  In the current calculation of
the stellar structure, however,  we ignore such a  heat transfer, for simplicity of the calculation.  This would be reasonable approximation because
the timescale of radiative transfer is of the order of  $\tau_r\sim R_{c}^2\rho_c\kappa /c\sim 10^5~(R_c/3\times 10^8~{\rm cm})^2 (\rho_c/10^8~{\rm g~cm^{-3}})(\kappa/10^{-2}{\rm cm^{2}~g^{-1}})~{\rm years}$, which is longer than the spin-down timescale when the MHD wind exists and the cooling timescale with  the initial radiation luminosity.

For the massive merger product ($M_{WD}>1.2M_{\odot}$), the temperature and the mass density around the core/envelope boundary may become high enough to ignite the Carbon burning process ($\rho>3\times 10^5~{\rm g~cm^{-3}}$ and $T>8\times 10^8$~K, respectively). This Carbon burning process can last about $10^{3-4}$ years and may leave an ONeMg WD. Such a burning process will prolong the cooling timescale and may enhance the spin-down of the merger product  due to a MHD wind. In the current study, we will mainly discuss the case in which the internal temperature does not reach to the threshold of the Carbon burning process.  We will discuss the influence  of the burning process on our results in section~\ref{sec:4-1}.

\subsubsection{Hydrostatic equations}
\label{hstru}
We solve  the structure of the merger product under the assumption of hydrostatic equilibrium,
\begin{equation}
  \frac{dP}{dr}=-\frac{Gm\rho}{r^2}, 
\end{equation}
with equation of stellar mass,
\begin{equation}
  \frac{dm}{dr}=4\pi r^2\rho.
\end{equation}
and the equation of state of
\begin{equation}
  P=P_i+P_{r}+P_d
\end{equation}
where $P_i=\rho k_BT/(\mu m_p)$ is the ideal gas law, $P_{r}=aT^4/3$ is the radiation pressure,  $P_d$ is the degenerate
pressure of the electron~\citep{1961ApJ...134..669S}, and  $a=4\sigma_{SB}/c$ with $\sigma_{SB}$ being the Stefan-Boltzmann constant. For the temperature profile,
we assume a constant temperature inside the  core and we apply
the core temperature of $T_c\sim 10^{7-8}$~K. For the envelope, we solve the energy transfer of
\begin{equation}
\frac{dT}{dr}=-\frac{3\kappa \rho L}{16\pi acr^2 T^3}.
\label{eq:trans}
\end{equation}
We assume the  luminosity profile of the  envelope as 
\begin{equation}
\frac{dL}{dM}={\rm constant,}
  \label{eq:lumi}
\end{equation}
with the condition that $L(M_{WD})=L_{WD}$. We ignore the radiation luminosity from the core, which with $T_{c}\sim 10^8$~K will be  much smaller
than $\sim 10^{37-38}~{\rm erg~s^{-1}}$ of the hot envelope.

\subsubsection{Boundary condition and solving procedure}
The model parameters to obtain a structure is the core temperature $(T_c$),
stellar mass ($M_{WD}$), envelope mass ($M_{env}$) and
stellar luminosity  ($L_{WD}$). For the mass, we impose the condition that 
\begin{equation}
m(0)=0~{\rm and}~m(R_{WD})=M_{WD},
\end{equation}
at the centre ($r=0$) and the outer boundary ($r=R_{WD}$), respectively.
We set the radius of the core/envelop boundary  ($r=R_c$) when the stellar 
mass integrated from the centre is equal to the core mass,  $M_{WD}-M_{env}$. For the radiation luminosity, we impose
the conditions of 
\begin{equation}
L(R_c)=0~{\rm and}~L(R_{WD})=L_{WD}.
\end{equation}
For the outer boundary, we determine the position by imposing the Eddington approximation of 
\begin{equation}
  T(R_{WD})=\frac{1}{2}T_{eff}~{\rm and}~P(R_{WD})\sim P_r(R_{WD}),
\end{equation}
where $T_{eff}=[L_{WD}/(4\pi\sigma_{SB} R^2_{WD})]^{1/4}$ is the effective temperature.

The process to obtain a structure of the merger product is as follows:
\begin{enumerate}
\item  We assume the central density
  and solve the structure of the core from the centre. Then we  find the radius of the core/envelope
  boundary, where $m(R_c)=M_{WD}-M_{env}$.

\item We assume the position of the outer boundary ($R_{WD}$) of the envelope
and solve the envelope structure from outer boundary to the core/envelope
boundary. When the pressure does not match to the core pressure at the core/envelope boundary,
we modify the position of the outer boundary.

\item  If the assumed position of the outer boundary satisfies  the pressure
  continuous  at the core/envelop boundary, we check the total mass of the envelope,
$M_{env}$. When the mass does not match to the desired mass of the envelop, we modify the central density of the core.  
\item We repeat above processes until all imposed conditions are satisfied.
  \end{enumerate}
We evaluate the condition of the slow-point with the obtained hydrostatic structure. Once the condition of the slow-point is satisfied at the envelope, we evaluate the angular momentum loss due to the MHD wind,  as described in section~\ref{sec:2.1}.

\section{Results}
\label{sec:3}
\subsection{Hydrostatic structure}
\label{sec:3.1}
Figure~\ref{fig:in} summarizes the properties of the merger product with the radiation luminosity of $L_{WD}\sim L_{\rm max}\sim 4\pi c GM_{WD}/\kappa_{\rm max}$
and its dependence on the core temperature: critical luminosity (tot left), central mass density (tot right), radius (bottom-left), and temperature at core/envelope boundary (bottom-right).  The figure also shows the dependency on the  mass of the product; the envelope is assumed
to be $M_{env}=0.3M_{\odot}$. From top-left panel, we can see that the critical luminosity, above which the energy transfer due to the convection becomes important in the envelope, is of
the order of $L_{\rm max}\sim (1-4)\times 10^4L_{\odot}$.  The figure does not display the results of $T_c>10^8$~K for $M_{WD}=1.0M_{\odot}$ and
$T_c>2\times 10^8$K for  $M_{WD}=1.1M_{\odot}$, because the radius ($R_{WD}$) exponentially increases above those central temperatures, as shown in
the bottom-left panel. Figure~\ref{fig:structure} shows the profiles of the pressure (left panel) and temperature (right panel) for $L_{WD}\sim L_{\rm max}$;
the solid and dashed lines are the results for $(M_{WD},T_c)=(1.3M_{\odot}, 3\times 10^8~{\rm K})$ and $(1.0M_{\odot}, 10^8~{\rm K})$, respectively. As we
described in section~\ref{sec:assumption}, we ignored the heat transfer from the envelop to core regions and hence
the temperature and mass density profiles have a discontinuity at the core/envelope boundary.

 To examine the stability against convection in the envelope, Figure~\ref{fig:ga} evaluate the Schwarzschild criterion of the convection in the envelop for the case of
$M_{WD}=1.3M_{\odot}$ in Figure~\ref{fig:structure}. The index $\gamma$ (solid line) and the adiabatic index $\gamma_a$ (dashed line) are  defined by
\begin{equation}
  \gamma\equiv \frac{\rho}{P}\left(\frac{dP}{d\rho}\right)_{\rm star},
  \end{equation}
and

\begin{equation}
 \gamma_a=\frac{32-24\beta-3\beta^2}{24-21\beta},
  \label{ga}
\end{equation}
respectively, where $\beta=P_{i}/P$ and the  degenerate pressure in the outer envelope is neglected. We can see in Figure ~\ref {fig:ga} that  the current calculation
with $L_{WD}\sim 4\pi cGM_{WD}/\kappa_{max}\sim 3\times 10^{4}L_{\odot}$ would predict a
convection motion around $r\sim 1.4\times 10^{10}$~cm and $\sim 1.55\times 10^{10}$~cm, where $\gamma>\gamma_a$. However, the majority  of the envelop satisfies   the condition $\gamma<\gamma_a$, indicating the  stability against convection. Hence, the current results  discussed  with
the radiative heat transfer of equation~(\ref{eq:trans}) may be validated.

\subsection{Slow-point at the envelope}
\label{sec:slowpoint}
As discussed in section~\ref{sec:2.1}, the stellar wind will be launched if the condition of the slow-point
expressed by equation~(\ref{eq:slow}) is satisfied in the envelope of the merger product.
The left panel of Figure~\ref{fig:opa} illustrates  the profile of the opacity (solid line) and function $F(r)$ (dashed line), which is defined by equation of (\ref{eq:func}),
at the outer envelope; in the figure we assume the angular speed of $\Omega_{WD}=0.5\Omega_{K}$ . The figure shows that the condition for the slow-point, where
$F(r) = 0 $, is met at two locations, reflecting the profile of the opacity near the stellar surface.  In this study, we consider that the slow-point not only satisfies
$F(R_s)=0$ but also $dF/dr|_{R_s}>0$. The latter condition may ensure  that
beyond the slow-point, the wind will be accelerated and the dynamics pressure comes to play an important role. For Figure~\ref{fig:opa}, for example, the slow-point is located at
$R_s\sim 1.45\times 10^{10}$cm, where $F(R_s)=0$ and $dF/dr|_{R_s}>0$ are satisfied, simultaneously.

It will be important to note that in the current model, the slow-point in the envelope appears as a result of  the effect of the spinning of the remnant product. For instance, there  is no solution satisfies the condition of the slow-point when $\Omega_{WD}=0$~s, as illustrated by the dotted line in the left panel of Figure~\ref{fig:opa}.
With the large radiation luminosity of $L_{WD}\sim 10^{4}L_{\odot}$, we can see that the influence of the gas pressure on the function $F(r)$,
which are the second term in the right hand side of equation~(\ref{eq:func}), is negligible. The solid line in the right panel of Figure~\ref{fig:opa} presents the
mass density - temperature relation at the outer envelope. It is found that with this profile of the mass density and temperature, the gas pressure term
in equation~(\ref{eq:func}) can be estimated about one or two order of magnitude smaller than the radiation pressure term. Consequently, the condition of the slow-point is approximately
described as
\begin{equation}                                                                                \frac{\kappa L(R_s)}{4\pi R_sc}-\frac{Gm(R_s)}{R_s}+R_s^2\Omega_{WD}^2\sim 0.                   \label{eq:slow1}
\end{equation}   
 As described in section~\ref{sec:assumption},  our solution of the hydrostatic structure is obtained under the condition of the stellar luminosity that $L_{WD}< L_{\rm max}$, above which
the convection describes the heat transfer.
This condition in the envelope is equivalent to  the condition that $\kappa L_{WD}/(4\pi rc)<GM_{WD}/r$. In the current model, hence, if there is no influence of the spin,
the function $F(r)$ takes always a negative value, $F(r)\le 0$,
in the envelope, and no stellar  wind is launched. 

The dashed line in the right panel of Figure~\ref{fig:opa} presents the opacity at the outer envelope divided by the Kramer formula of
\begin{equation}
  \kappa_{K}\sim \kappa_0\rho T^{-7/2},
  \label{eq:kramer}
\end{equation}
where $\kappa_0\sim 4\times 10^{25}$ in standard C.G.S. units. We can see that the ratio $\kappa/\kappa_{K}$ is about unity  at the outer envelope, except  for near the outer boundary. This indicates that the opacity at the outer envelope is generally  described by the Kramer formula of equation~(\ref{eq:kramer}). In section~\ref{sec:estimation},  we will apply the Kramer formula to estimate the mass-loss rate of the MHD wind.

\subsection{Cooling timescale and spin-down timescale}
\label{sec:estimation}
To investigate  the temporal evolution of the merger product, we evaluate  the cooling process of the merger product.  When the nuclear burning process at the core/envelope boundary could be avoided, the merger product will evolve with a Kelvin-Helmholtz timescale.
In the calculation, therefore, we carry out the grid calculation for the luminosity $L_{WD}$. For example, we prepare 200 grids for the luminosity in the range of $10^3L_{\odot}~\le L_{WD} \le L_{\rm max}$. Figure~\ref{fig:grav} show the energy component of the merger-product as a function of the radiation luminosity for the case of $(M_{WD}, M_{env}, T_c)=(1.3M_{\odot}, 0.3M_{\odot}, 3\times 10^8$K); $\Phi_G$ is the gravitational potential energy, $U_i$, $U_r$ and $U_d$ represent the internal energies of the ideal gas, radiation and degenerate gas, respectively.
We evaluate the evolution timescale from   
\begin{equation}
  \triangle t_i=\frac{|\triangle\Phi_G+\triangle U_i+\triangle U_r+\triangle U_d |}
            {L_{WD}+L_{\nu}},
            \label{eq:evtimescale}
\end{equation}
where  $L_{\nu}$ represents the luminosity carried by the neutrino~\citep{1989ApJ...339..354I}.

Figure~\ref{fig:cool} summarizes the temporal evolution of the structures of the merger
product; radiation luminosity (top-left), radius (top-right), the mass density (bottom-left)
and temperature (bottom-right) at the core/envelope boundary,
respectively. For the parameters of $(M_{WD},T_c)=(1.3M_{\odot}, 3\times 10^8~{\rm K})$ (solid lines)  and $(1.0M_{\odot}, 10^8~{\rm K})$ (dashed lines),
the evolution timescale of the initial stage is of the order
of $10^{3-4}$~years. As the bottom-right panel shows, the temperature at the core/envelope boundary initially increases with time as a result of the compression of the envelope. It reaches
maximum temperature at $\sim 10^{4}$~years. For $M_{WD}=1.3M_{\odot}$ (solid line), we find that if the core temperature of the merger product
is $T_c\sim 3\times 10^8$~K, the maximum temperature inside does not reach to the threshold, $\sim 8\times 10^8$~K,  of the Carbon burning process, and it could be possible to produce
a massive CO WD. For a core temperature of $T_c<2\times 10^8$~K, on the other hand, the maximum temperature can exceed the critical temperature of the Carbon burning, as Figure~\ref{fig:structure} indicates. The  structure such as the peak temperature at the core/envelope boundary
is roughly consistent with the previous studies \citep[e.g.,][]{shen2012long}.

Although we numerically solve the position of the slow-point and mass-loss rate of the MHD
wind, we can perform a semi-analytical investigation for them.  
As discussed in section~\ref{sec:slowpoint}, the condition of the slow-point in the current model of the merger-product  can be approximated by equation~(\ref{eq:slow1}). We define $\omega$ to be the ratio
\begin{equation}
  \omega\equiv \frac{\Omega_{WD}}{\Omega_K}.
\label{eq:omega}
\end{equation}
As illustrated in Figure~\ref{fig:opa},  the distance between the slow-point and the outer boundary is much shorter than the radius of the merger product. We 
therefore assume $R_s\sim R_{WD}$, $L(R_s)\sim L_{WD}$ and $m(R_s)\sim M_{WD}$, Using equations~(\ref{eq:slow1}) and~(\ref{eq:omega}), the opacity at the slow-point is estimated as 
\begin{eqnarray}
  \kappa(R_s)&\sim& 4\pi (1-{\omega^2})\frac{GM_{WD}c}{L_{WD}} \nonumber\\
  &\sim& 0.5(1-{\omega^2})\left(\frac{M_{WD}}{M_{\odot}}\right)
  \left(\frac{L_{WD}}{10^{38}~{\rm erg~s^{-1}}}\right)^{-1}{\rm cm^{2}~g^{-1}}.
\label{eq:kappas}
\end{eqnarray}

To estimate the mass density at the slow-point, we apply the Kramer formula of equation~(\ref{eq:kramer}). In addition, we  introduce the parameter $f_T$, and represent the temperature at the slow-point as $T(R_s)=f_TT_{eff}=f_T[L_{WD}/(4\pi\sigma_{SB} R^2_{WD})]^{1/4}$. From  Figure~\ref{fig:opa}, we can see that the parameter, $f_T$, is a factor of several. Consequently, the mass density at the slow-point is estimated as 
\begin{eqnarray}
  \rho(R_s)&\sim& \frac{\kappa(R_s)}{\kappa_0}f_T^{7/2}T_{eff}^{7/2}
\sim  5\times 10^{-7}\left(\frac{f_T}{2}\right)^{7/2}\left(\frac{\kappa}{0.5~{\rm cm^2~g^{-1}}}\right) 
  \nonumber \\
 &&
  \left(\frac{L_{WD}}{10^{38}{\rm erg~s^{-1}}}\right)^{7/8}
\left(\frac{R_{WD}}{10^{10}~{\rm cm}}\right)^{-7/4}~{\rm g~cm^{-3}},
\label{eq:rhos}
\end{eqnarray}
where we scale the parameter $f_T$ using the typical value of our results.
Inserting equation~(\ref{eq:kappas}) into equation~(\ref{eq:rhos}) and using 
equation~(\ref{eq:massloss0}), the mass-loss rate is estimated to be 
\begin{eqnarray}
  \dot{M}_W&= &4\pi R_s^2\rho(R_s)c_s
  \sim f_T^4\frac{ \kappa L_{WD}}{\kappa_0 \sigma_{SB}}\sqrt{\frac{k_B}{\mu m_p}} \nonumber \\
  &\sim& 10^{21}
  (1-{\omega^2})\left(\frac{f_T}{2}\right)^{4} \left(\frac{M_{WD}}{M_{\odot}}\right)~~{\rm g~s^{-1}} .
  \label{eq:massloss}
\end{eqnarray}
The current model, therefore, predicts that
 if the MHD wind from the merger product exists,  
 its  mass-loss rate is roughly of the order of $\dot{M}_W\sim 10^{20-21}~\rm {g~s^{-1}}$. The resultant spin-down will happen  with a timescale  of
 \begin{eqnarray}
   \tau_w&\sim& \frac{J_{WD}}{dJ_{WD}/dt}\sim 
   80 \omega^{1/3}\left(\frac{I}{10^{51}{\rm g~cm^2}}\right)
   \left(\frac{M_{WD}}{M_{\odot}}\right)^{1/3}\left(\frac{R_{WD}}{10^{10}~{\rm cm}}\right)^{-1} \nonumber \\
     &\times& \left(\frac{\dot{M}_W}{10^{21}~{\rm g~s^{-1}}}\right)^{-1/3}
     \left(\frac{\Phi_B}{10^{25}~{\rm G~ cm^2}}\right)^{-4/3}~{\rm years}.
\label{eq:timescale}
 \end{eqnarray}

\subsection{Spin-down process due to MHD wind}

\begin{figure*}
  \includegraphics[scale=0.9]{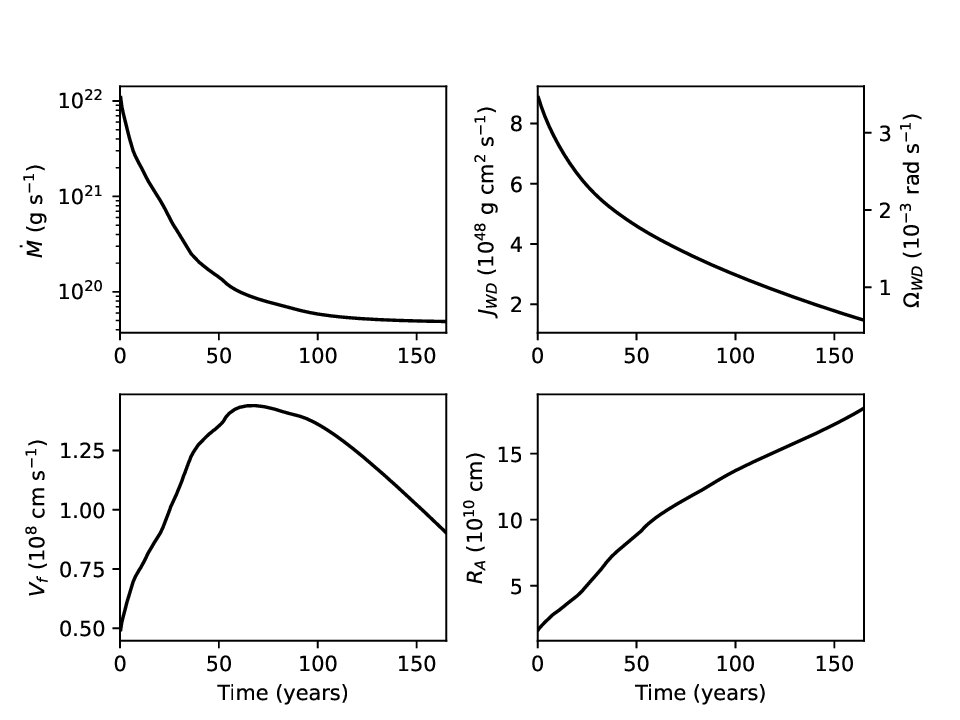}
  \caption{ Example of the temporal evolutions of the wind properties and spin of the merger  product with the mass of $M_{WD}=1.3M_{\odot}$.
    Top-left: Wind mass-loss rate. Top-right: Angular moment and spin frequency. Bottom-left: Terminal speed of the wind. Bottom-right:
    Alf\'{v}en radius. The magnetic flux is $\Phi_B=10^{25}{\rm G~cm^2}$ and the other parameters of the merger products are same as those in Figure~\ref{fig:cool}.}
  \label{fig:wind}
\end{figure*}

\subsubsection{Temporal evolution of wind properties}
Figure~\ref{fig:wind} presents an example of the properties of the MHD wind from the merger product with $M_{WD}=1.3M_{\odot}$, $M_{env}=0.3M_{\odot}$, $T_c=3\times 10^8$~K  and the initial radiation luminosity of $L_{WD}=3\times 10^4L_{\odot}$, for which the cooling evolution  is presented in Figure~\ref{fig:cool} (solid line). We assume the initial angular speed and the magnetic flux of  $\Omega_{WD}=0.5\Omega_K\sim 3.3\times 10^{-3}{\rm rad~s^{-1}}$ ($J_{WD}=8.5\times 10^{48}{\rm g~cm^{2}~s^{-1}}$) and $\Phi_B=10^{25}{\rm G~cm^2}$, respectively. As shown in the figure, the angular momentum decreases with a timescale of a hundred year, which can be explained by equation of (\ref{eq:timescale}). The launch of the MHD wind is terminated when $\Omega_{WD}\sim 6.4\times 10^{-4}{\rm rad~s^{-1}}$ at $t\sim 160$~years, as top-right panel indicates.

As shown in top-left panel in Figure~\ref{fig:wind}, the initial wind mass-loss rate is
of the order of $\dot{M}_W\sim 10^{22}~{\rm g~s^{-1}}$, and the mass-loss rate decreases with time.
The decrease in the mass-loss rate is explained as follows. The initial profile of the function $F(r)$ is, for example, described by the dashed line in the left panel of
Figure~\ref{fig:opa}, in which the slow-point (filled circle) is located at $R_s\sim 1.45\times 10^{10}$~cm with  $f_T\sim 2.5$. As the spin-down timescale due to the MHD wind described by equation~(\ref{eq:timescale}) is usually shorter than or comparable to the cooling timescale, the structure of the
merger product does not significantly change during the spin-down.  This indicates
that as spinning down, the profile of the function $F(r)$ in Figure~\ref{fig:opa}  vertically descends toward dotted-line for the case of  $\Omega_{WD}=0$~s, and the position of the slow-point
shifts outward. This  reduces  the parameter $f_T$ from  2-3 to nearly  unity.  Consequently, just before the wind is terminated, the mass-loss rate is
  of the order of $\sim 5\times 10^{19}~{\rm g~s^{-1}}$, as indicated in the equation~(\ref{eq:massloss}) and Figure~\ref{fig:wind}.

\subsubsection{Final angular momentum}
In our model, the final angular frequency is determined as follows. As Figure~\ref{fig:opa} shows, the opacity reaches the maximum value $(\equiv \kappa_{\rm max})$ in the envelope. From equation~(\ref{eq:slow1}) of the condition of the slow-point, the final angular frequency is approximated by
\begin{equation}
  \Omega_{f}\sim \Omega_K\left(1-\frac{\kappa_{\rm max}L_{WD}}{4\pi cGM_{WD}}\right)^{1/2}.
\label{eq:final}
\end{equation}
In the case of Figure~\ref{fig:wind}, for example, the initial  cooling timescale ($\sim 10^3$ years, Figure~\ref{fig:cool}) is longer than the spin-down timescale. As a result, the final angular frequency (angular momentum) is determined by the initial radiation  luminosity. For the case of  Figure~\ref{fig:wind}, when the stellar wind
is terminated, the merger remnant has the angular momentum of $J_{WD}\sim 1.6\times 10^{48}~{\rm g~cm^2~s^{-1}}$ and it is still in the giant phase with the
moment of inertia of $\sim 2.6\times 10^{51}~{\rm g~cm^{2}}$. During the contraction of the envelope, the angular momentum is conserved but the angular frequency increases
because of the reduce of the moment of inertia.  After enough cooling process, the merger remnant with $M_{WD}=1.3~M_{\odot}$ will
become a WD with a momentum of inertial of $\sim 4.5\times 10^{49}~{\rm g~cm^{2}}$, indicating the angular frequency of the WD becomes
$\Omega_{WD}\sim 0.034~{\rm rad~s^{-1}}$ or spin period of $P_{WD}\sim 185$~s if we ignore the spin-down due to the magnetic dipole radiation.

\begin{figure}
  \includegraphics[scale=0.5]{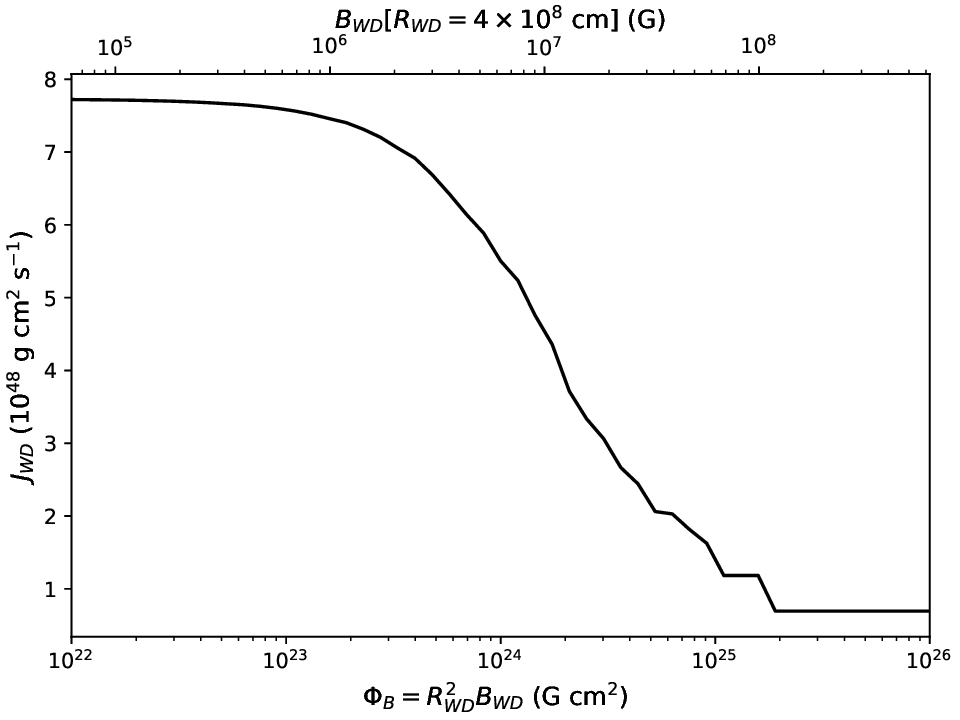}
  \caption{ The final angular momentum  when the wind is terminated, expressed as a function of the stellar magnetic flux.
    The expected surface magnetic field after enough cooling, for which the radius $R_{WD}=4\times 10^8$~cm is used, is also indicated in
    upper axis. The total mass and envelope mass are $M_{WD}=1.3M_{\odot}$ and $M_{env}=0.3M_{\odot}$, respectively. The initial
    luminosity and angular frequency are
    $L_{WD}\sim 3\times 10^4L_{\odot}$ and $\Omega=0.5\Omega_K$, respectively.}
  \label{fig:fbm}
\end{figure}

Figure~\ref{fig:fbm} shows the dependency of the final angular momentum as a function of the magnetic flux; the parameters of $M_{WD}$, $M_{env}$, $T_c$
and the initial radiation luminosity $L_{WD}=3\times 10^4L_{\odot}$ are the same as the cases of Figures~\ref{fig:structure}-\ref{fig:wind}. In the
current model, we divide the spin-down evolution into three types depending on the magnetic filed; (i) strongly magnetized case, (ii) mildly magnetized case and (iii) weakly magnetized case. For the strongly magnetized case, $\Phi_B\ge 7\times 10^{24}~{\rm G~cm^{2}}$ in Figure~\ref{fig:fbm},
the final angular momentum is independent  on the magnitude of the magnetic field strength. This is because the  initial spin-down timescale is shorter than
the cooling timescale with the initial radiation luminosity. As a result, the final angular speed is approximately  determined by equation~(\ref{eq:final}) with the initial radiation luminosity. For the mildly magnetized case, $10^{23}~{\rm G~cm^{2}} \le\Phi_B\le 3\times10^{24}~{\rm G~cm^{2}}$,
the initial spin-down timescale is longer than  the initial cooling timescale, and the merger product can be cooled down during the MHD wind exists.
The MHD wind will be terminated when the cooling timescale is comparable to the spin-down timescale.
For weakly magnetized   case, $\Phi_B<10^{23}~{\rm G~cm^{2}}$, the  spin-down timescale is too long and the MHD wind is ineffective to the
spin-down of  the merger product.

For the case that the initial angular speed and/or the initial luminosity are too small  so that $F(r)<0$ is satisfied over envelope, no wind from the merger product
is launched and no spin-down happens.  In such a case, the current observed spin period of the magnetic WD will be determined by initial angular momentum or by other spin-down process.  Our model also implies that when the initial luminosity is very close to  $L_{\rm max}\sim4\pi c GM_{WD}/\kappa_{\rm max}$,
the MHD wind  can carry away a significant fraction of the  initial angular momentum and produce a slowly rotating WD (e.g. PG~1031+234).

\subsection{Application}
\begin{figure}
  \includegraphics[scale=0.5]{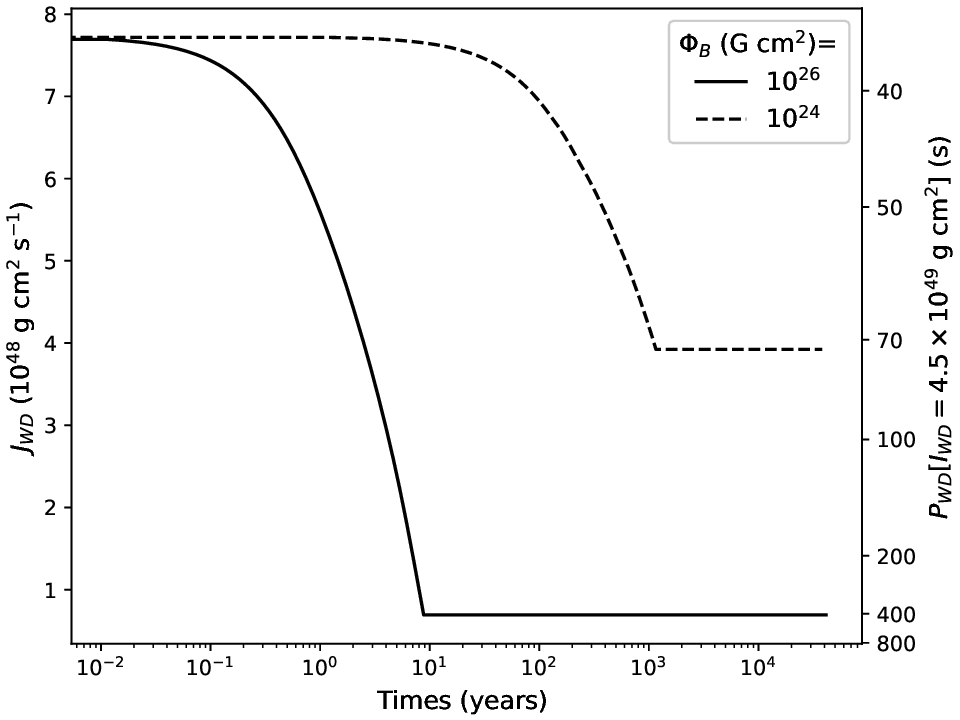}
  \caption{ Model of the temporal evolution of the angular momentum for J1901+1458 (solid line) and J2211+1136 (dashed line). The results are for $M_{WD}=1.3M_{\odot}$, $M_{env}=0.3M_{\odot}$, $T_{c}=3\times 10^8$~K and the initial radiation luminosity $L_{WD}\sim 3\times 10^4L_{\odot}$. The expected spin period $P_{WD}$ once the merger remnant is sufficiently cooled down is also indicated using the right vertical
  axis.}
  \label{fig:appli1}
\end{figure}

\subsubsection{J1901+1458 and J2211+1136}
\label{sec:application}
In this section, we apply the model to two WDs, J1901+1458 and J2211+1136, which have a current spin period of $P_{WD}\sim 416.4$~s
and $\sim 70.32$~s, respectively, and we  assume  that the current angular momentum was mainly determined by the spin-down as a result of the
MHD wind.  Since the measured masses of J1901+1458 and J2211+1136 are $M_{WD}\sim 1.327-1.365M_{\odot}$
and $1.27M_{\odot}$, respectively, we apply  the MHD wind of the case of $1.3M_{\odot}$. Using the degenerate pressure of the zero temperature limit
~\citep{1961ApJ...134..669S}, the moment of inertia of the WD with $1.3M_{\odot}$ is $I_{WD}\sim 4.5\times 10^{49}~{\rm g~cm^{2}}$, which
implying current angular momenta of J1901+1458 and J2211+1136 are $J_{WD}\sim 6.8\times 10^{47}~{\rm g~cm^{2}~s^{-1}}$ and $4.0\times 10^{48}~{\rm g~cm^{2}~s^{-1}}$, respectively. Our model predicts that the current angular momentum has remained nearly constant after the wind
ceased, as the angular momentum loss due to magnetic dipole radiation is negligible.
The measured magnitudes of the magnetic field of J1901+1458  and   J2211+1136
are $B_{WD}\sim (6-9)\times 10^8$~G and $\sim 1.5\times 10^7$~G, respectively. The radius of the WD with $M_{WD}\sim 1.3M_{\odot}$ is
$R_{WD}\sim (3-4)\times 10^{8}$~cm, suggesting J1901+1458  and   J2211+1136 have
a magnetic flux of the order of  $\Phi_{B}\sim 10^{26}~{\rm G~cm^2}$ and $\sim 10^{24}~{\rm G~cm^2}$, respectively.

Figure~\ref{fig:appli1} shows an example of the temporal evolution of the angular momentum under the effect of the MHD wind
to reach the observed values for  J1901+1458 (solid line) and J2211+1136 (dashed line): the model parameters are
the initial luminosity of $L_{WD}\sim 3\times 10^4L_{\odot}$, the core temperature of  $T_c=3\times 10^8$~K and envelope mass of 
$M_{env}=0.3M_{\odot}$. The calculation  is also taken into account the spin evolution due to  the magnetic dipole radiation after the termination of the MHD wind.

As shown in the figure,  the angular momenta of two cases are decreased with a timescale of $\sim 10$~years for J1901+1458 (solid line) and $\sim 10^3$~years for J2211+1136 (dashed line), respectively. This timescale can be explained by equation~(\ref{eq:timescale}).
For J1901+1458, the spin-down timescale with the initial angular frequency  is
shorter than the cooling timescale with the initial luminosity of 
$L_{WD}\sim 3\times 10^4L_{\odot}$. In such a case, the angular frequency when the wind is terminated is determined by
equation~(\ref{eq:final}) with the initial luminosity.  For  J2211+1136, on the other hand, the spin-down timescale
is longer than the initial cooling timescale, and the spin-down process is influenced by the cooling process of the merger product. In such a case, the final angular frequency is determined by the equation~(\ref{eq:final}) with the luminosity, at which the spin-down timescale is equal to the cooling timescale. After the wind is terminated, the angular momentum is almost constant with time since the timescale of 
the spin-down as a result of the magnetic dipole radiation (equation~(\ref{eq:dipole})) is much longer than the cooling timescale.

\begin{figure}
  \includegraphics[scale=0.5]{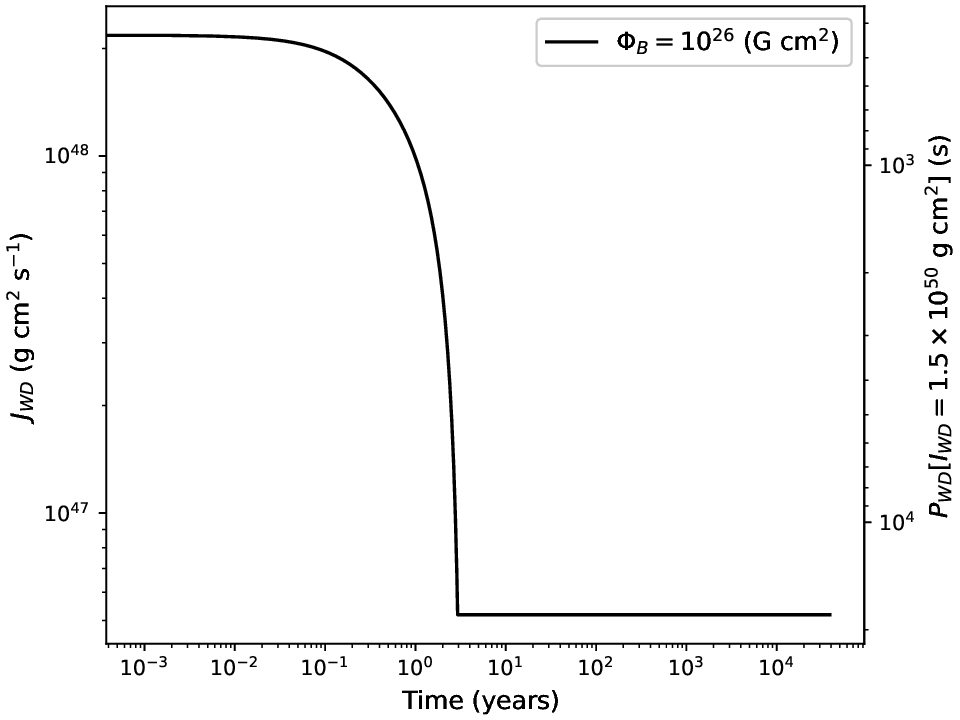}
  \caption{ A spin-down model for PG~1031+234 with $\Phi_{B}=10^{26}~{\rm G~cm^2}$. The parameters are $M_{WD}=M_{\odot}$, $M_{env}=0.2M_{env}$, $T_c=10^8$~K, initial angular frequency $\Omega_{WD}=0.5\Omega_K$ and initial radiation luminosity
  $L_{WD}\sim 2.2\times 10^{4}L_{\odot}$. The expected spin period after the sufficiently cooing down is indicated using the right vertical axis.}
  \label{fig:appli2}
\end{figure}

\subsubsection{PG 1031+234}
PG~1031+234 is a classical magnetic WD with a magnetic field of $B_{WD}\sim (2-10)\times 10^8$~G and it has  a smaller mass ($M_{WD}\sim M_{\odot}$) and a longer-spin period
($P_{WD}\sim 212$~min) compared to those of J1901+1458 and J2211+1136.  Using the degenerate pressure of the zero temperature limit,
one solar mass WD  has a radius of $R_{WD}\sim5\times 10^8$~cm and moment of inertia of $I_{WD}=1.3\times 10^{50}~{\rm g~cm^2}$,
which corresponds to the magnetic flux of $\Phi_B\sim 10^{26}~{\rm G~cm^2}$ and current angular momentum of $J_{WD}\sim 6.4\times 10^{46}~{\rm g~cm^2~s^{-1}}$. This angular momentum is significantly smaller than  those of   J1901+1458  and J2211+1136 discussed in section~\ref{sec:application}.  Hence, if PG 1031+234 resulted from a double WD merger and its initial angular frequency was approximately Keplerian value, it experienced a stronger spin-down as a result of the MHD wind.

Figure~\ref{fig:appli2} shows the model of the temporal evolution of PG 1031+234; the parameters are $M_{WD}=M_{\odot}$,  $M_{env}=0.2M_{\odot}$, $T_c=10^8$~K, initial angular frequency $\Omega_{WD}=0.5\Omega_K$ and initial radiation luminosity  $L_{WD}\sim 2.2\times 10^{4}L_{\odot}$. Similar to J1901+1458, the
initial spin-down timescale due to the MHD wind is much shorter than the cooling timescale that is in the order of $\sim 10^3$~years. Hence the final frequency of equation~(\ref{eq:final}) is
determined by the initial radiation luminosity. As indicated by equation~(\ref{eq:final}), when the radiation luminosity is very close to the Eddington  
value of $L_{WD}\sim 4\pi c GM_{WD}/\kappa_{\rm max}$, which is $\sim 2.2\times 10^{4}L_{\odot}$ in the current  model of the stellar structure, the final angular frequency becomes significantly smaller than
the Keplerian value.  Our model therefore expects that PG 1031+234 appeared with 
the  initial luminosity that is almost the Eddington value. 

As indicated by the function form of equation~(\ref{eq:final}), the final angular frequency  is sensitive to the initial luminosity when the luminosity is close to  $L_{\max}$. Two magnetic WDs, J1901+1458 ($P_{WD}\sim 416.4$~s) and PG~ 1031+234 ($P_{WD}\sim 1.3\times 10^4$~s),  may have similar magnetic field strength ($B_{WD}\sim (5-9)\times 10^8$~G) but appear with very different spin periods.  
The sensitivity of the final angular frequency on the  initial luminosity would explain the 
 difference in the spin periods of the two WDs.

\section{Discussion}
\label{sec:4}
\subsection{Influence on shell nuclear burning process}
\label{sec:4-1}
\begin{figure}
  \includegraphics[scale=0.5]{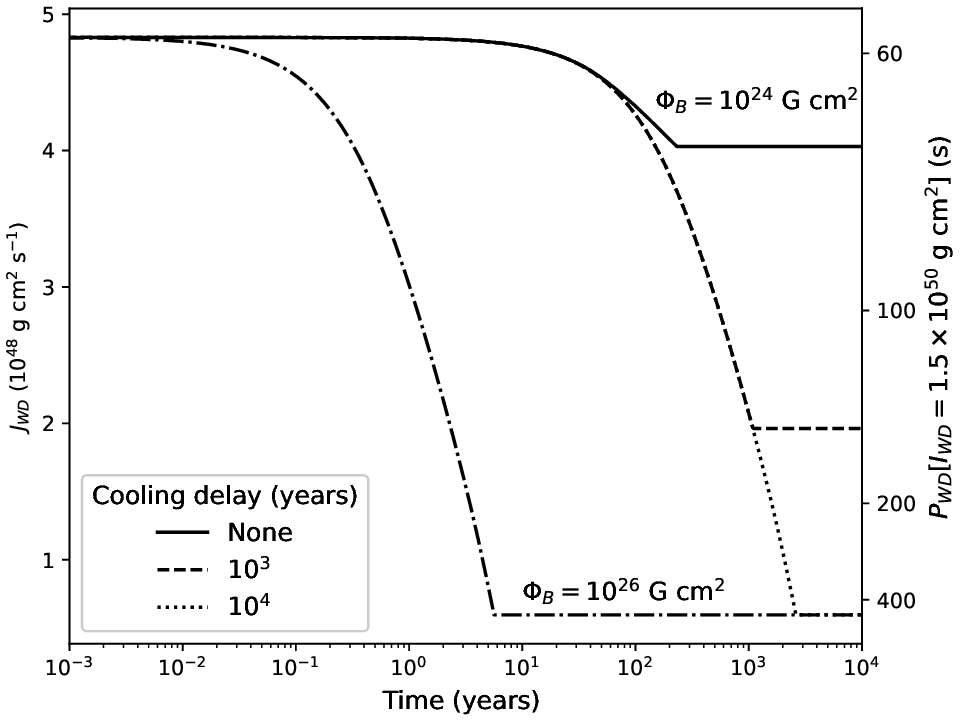}
  \caption{ Influence of the cooling delay due to the shell-burning on the spin-down evolution. The dashed-dotted line is the case of $\Phi_{B}=10^{26}~{\rm G~cm^2}$, for which the cooling delay does not influence the final angular momentum. 
    The solid line is the cases for $\Phi_{B}=10^{24}~{\rm G~cm^2}$ without cooling delay, and
    the dashed and dotted lines are the case of a  cooling delay of $10^3$~years and $10^4$~years, respectively. The parameters are
    $M_{WD}=1.3M_{\odot}$, $M_{env}=0.3M_{\odot}$, $T_c=2\times 10^8$~K and
    initial luminosity of $L_{WD}\sim 3.5\times 10^4L_{\odot}$.}
  \label{fig:burning}
  \end{figure}
The temperature inside the merger product reaches the peak at the core/envelope boundary,  as
shown in Figure~\ref{fig:structure}, and this boundary  temperature  depends on the core temperature,  as indicated in Figure~\ref{fig:in}. During the envelope contraction phase, moreover,  the boundary temperature reaches its maximum value  at the age of $t\sim 10^4$~years, as Figure~\ref{fig:cool} indicates. In section~\ref{sec:3}, we examined  the stellar structure with the core temperature of $T_c=3\times 10^8$~K for $M_{WD}=1.3M_{\odot}$ and $10^8$~K for $M_{WD}=M_{\odot}$.
This was because the temperature at the core/envelop boundary  during the envelope contraction  does not reach $\sim 8\times 10^8$~K of the threshold for Carbon  burning process. When the core temperature for $M_{WD}=1.3M_{\odot}$ is $T_c\le 2\times 10^8$~K, on the other hand, the boundary temperature will reach the threshold value and the  Carbon burning will be ignited, as indicated in
Figure~\ref{fig:in}.  It has been argued that the Carbon shell-burning will last about $10^{3-4}$~years~\citep{schwab2016evolution},
and  this shell-burning process could  delay the cooling process.

We speculate that the shell-burning process will influence the spin-down for the case
that the initial spin-down timescale is longer than the initial cooling timescale when the burning process is neglected; namely,  the weakly
and mildly magnetized cases for  $\Phi_B<5\times 10^{24}~{\rm G~cm^2}$
in Figure~\ref{fig:fbm}. Figure~\ref{fig:burning} represents a toy model for the influence of the burning process. We assume the burning process starts from the initial stage and we  artificially delay  the initial cooling timescale by  $10^3$~years or $10^4$~years with a constant radiation luminosity; we assume the core temperature of $T_c=2\times 10^8$~K,  $M_{WD}=1.3M_{\odot}$, $M_{env}=0.3M_{\odot}$ and the initial luminosity of $L_{WD}\sim 3.5\times 10^{4}L_{\odot}$, with which the boundary temperature can reach to $T_b\sim 8\times 10^8$~K of the threshold for Carbon  burning (Figure~\ref{fig:in}).

Figure~\ref{fig:burning} indicates that for the  strongly  magnetized product (the dashed-dotted line), the cooling  delay does not influence the final angular momentum.
This is because the initial spin-down timescale is $\sim 10$~years, which is much shorter than the cooling timescale.  For the mildly
magnetized case (solid, dashed
and dotted lines) in the figure, the initial spin-down timescale is of the order of $\sim 10^3$~years, and the cooling delay can
influence the final angular momentum. As indicated by the dotted-line in the figure, if the cooling delay as a result of the Carbon-burning process
is longer than the spin-down timescale, the final angular momentum is determined by the luminosity during the burning process.

Figure~\ref{fig:burning} assumes that the Carbon-burning process begins at the initial stage of the evolution. It will be possible, on the other hand, that  the  boundary temperature is initially too low   for the burning-process to start, but it reaches necessary  threshold due to the envelope contraction. In such a case, onset of the burning-process will be delayed by $10^{3-4}$~years, corresponding to the contraction timescale. We may expect that
if the timescale for the onset of the burning process is longer than the spin-down timescale due to  the MHD wind,
the burning process does not significantly  affect the spin-down evolution. Since the start time of the shell-burning process
and the changes in  the stellar structure and the radiation luminosity  as a result of  the burning process all affect the spin-down evolution,
a more sophisticated model will be  needed to discuss its influence on  the spin-down evolution of the merger product.

 \subsection{Comparison with previous works}

 The  evolution of the merger product  has mainly been discussed with two scenarios, namely, (i) a WD surrounded by the accretion
 disk~\citep{1985ApJ...297..531N,2007MNRAS.380..933Y,kulebi2013magnetic, sousa2022double}
 and (ii) a giant-star-like remnant~\citep{shen2012long, 2012MNRAS.427..190S, schwab2016evolution, wu2022formation}.  For the first scenario,  a WD with a hot envelope is surrounded by an accretion disk, which is the remnant of the disrupted secondary WD, and the phase of the accretion from the Keplerian disk with an Eddington limit will continue for a timescale of
 $\sim$ hours  to $\sim 10^5$~years. For the second scenario, the magnetic
 stress rapidly redistributes angular momentum from the core to Keplerian disk, producing a solid-rotating merger product without accompanying the accretion disk. \cite{shen2012long}
 argue that the timescale of the redistribution  is $\sim 10^4$~s, which is much shorter than the cooling timescale, leading  the merger product
 to enter the giant phase.  They also suggest  that the rotation energy of the Keplerian disk is converted into heat and the giant-star-like object
 appears as a near-Eddington source.

 Within the framework of the first scenario,  \cite{sousa2022double}
 discuss the spin-evolution for J1901+1458 and J2211+1136 under the effect of the debris disk and the dipole radiation.
 In their calculation, the accretion rate being  constant with time is assumed and the accretion from the disk continues
 until all initial mass of the debris disk is consumed.  In their model, the observed spin periods of the magnetic WDs almost represent the equilibrium spin periods in the accretion stage, in which the co-rotation radius is equal to the  Alf\'{v}en radius.
 Within the framework of the second scenario, we  discussed the spin evolution  as a result of   the MHD wind,
 without considering the effect of the accretion disk, and applied the model to J1901+1458, J2211+1136 and PG~1031+234.

Because of the theoretical uncertainties, we may consider a potential influence of the accretion disk if it would exist in the giant phase.  We assume  the accretion process with an  Eddington rate, which in the giant phase is of the order of 
\begin{eqnarray}
  \dot{M}_{E}&=&\frac{4\pi cR_{WD}}{\kappa} \nonumber \\
 & \sim& 10^{23} \left(\frac{R_{WD}}{5\cdot 10^{10}~\rm{cm}}\right)
  \left(\frac{\kappa}{0.2~{\rm cm^{2}~g^{-1}}}\right)^{-1} {\rm g~s^{-1}}.
\end{eqnarray}
We may express the torque exerted from the disk to the  WD as
\begin{equation}
  \dot{J}_{D}=\dot{M}_{E}R_{in}^2\Omega_K(R_{in})[1-\omega(R_{in})],
  \label{troque}
  \end{equation}
where  $R_{in}$ is the distance to the edge of the disk from the center of the WD, $\Omega_K(R_{in})=\sqrt{(GM)/R_{in}^3}$
is the Keplerian angular speed at the inner edge and $\omega(R_{in})\equiv\Omega_{WD}/\Omega_K(R_{in})$ is the so-called fastness parameter. When $\omega(R_{in})<1$, namely, the edge of the disk rotates faster than the WD, the disk matter accretes  on the  WD's surface and spins up the WD. When $\omega(R_{in})>1$, on the other hand, the accretion system is in a propeller regime and the WD is spun down.  The inner radius of the disk is determined by the equilibrium between the dynamics pressure of the disk and magnetic pressure of the WD, and it will be of the order of the  Alf\'{v}en radius of the accretion system, $R_{in}\sim \xi R_A\sim \xi [\Phi^2_BR_{WD}^2/(\dot{M}_E\sqrt{2GM})]^{2/7}$, where $\xi$ is a factor
of the order of unity. We can see that with $\dot{M}_E\sim 10^{23}~{\rm g~s^{-1}}$ and $\Phi_B<10^{26}~{\rm G~cm^2}$, the aforementioned inner radius, $R_{in}$, is smaller than $R_{WD}$, indicating the torque from the disk will exert at  the stellar surface, $R_{in}\sim R_{WD}$.  The ratio of the
torques from the accretion disk and from the MHD wind will be 
\begin{eqnarray}
  \frac{\dot{J}_{D}}{{\dot{J}_W}}&\sim& \frac{\dot{M}_{E}R_{WD}^2}{\dot{M}_{W}R^2_{A}\omega}\sim 25\omega^{-1} 
    \left(\frac{\dot{M}_E}{10^{23}~{\rm g~s^{-1}}}\right) \nonumber \\
&\times&  \left(\frac{\dot{M}_W}{10^{21}~{\rm g~s^{-1}}}\right)^{-1}
\left(\frac{R_{WD}}{5\cdot 10^{10}~{\rm cm}}\right)^2\left(\frac{R_{A}}{10^{11}~{\rm cm}}\right)^{-2},
\end{eqnarray}
where $\omega$ is given by equation~(\ref{eq:omega}). We find that during the accretion phase, the torque exerted from the accretion disk
can be  more impact on the spin evolution  than that from the MHD wind. We speculate
that the accretion process will keep the stellar angular frequency to be the  Keplerian value, namely, $\omega\sim 1$ and  will last about $M_{disk}/\dot{M}_E\sim 67 (M_{disk}/0.1M_{\odot})(\dot{M}_E/10^{23}~{\rm g~s^{-1}})^{-1}$~years, where $M_{disk}$ is the initial mass of the disk. This timescale of the accretion is shorter than
the cooling timescale of the merger product. After the accretion is stopped, therefore, the merger product will be spun-down by the
MHD wind, as discussed in section~\ref{sec:3}.

\cite{kulebi2013magnetic} discuss the evolution of the spin due to the stellar wind, based on the scenario that a
WD is  surrounded by the accretion disk (the first scenario). They also discuss the effect of the launch of the MHD wind on the spin evolution by assuming that the mass-loss rate is of the order of the mass accretion rate from the disk.
They suggest  that the accretion process and hence MHD wind   terminates once the disk temperature decreases a critical value, below which the turbulence in the disk is not sustained, and that the timescale of the disk survived is of the order of $10^5-10^7$~years. They  conclude that the launch of the wind enhances the spin-down of the WD.

In our scenario, the final angular momentum as a result of the MHD wind is determined by equation~(\ref{eq:final}) and it depends on the initial luminosity when
the merger product appears as a giant-star-like object. When the initial luminosity is $L_{WD}\ll 4\pi c GM_{WD}/\kappa_{\rm max}$, the wind will not be launched or
the influence of it on the spin-down will be negligible. In such a situation, the observed periods of the magnetic WDs were determined by
the initial angular momentum or by the different processes. When the merger product  is appeared with the initial luminosity
very close to $L_{WD}\sim 4\pi c GM_{WD}/\kappa_{\rm max}$, the MHD wind can carry away the significant fraction of the initial angular momentum.
As we have discussed, our calculation of the hydrostatic equilibrium with the radiative energy
transfer of equation~(\ref{eq:trans})  is  restricted  to the case of $L_{WD}<4\pi c GM_{WD}/\kappa_{\rm max}$. Since the current model suggests that  the
initial radiation luminosity is an important factor to characterize the final angular momentum.  a subsequent study that (i)   discusses the structure of the hot envelope with the convection energy transfer  and (ii) determination of the initial luminosity will be required.

\subsection{Limitation of model}
\begin{figure}
  \includegraphics[scale=0.5]{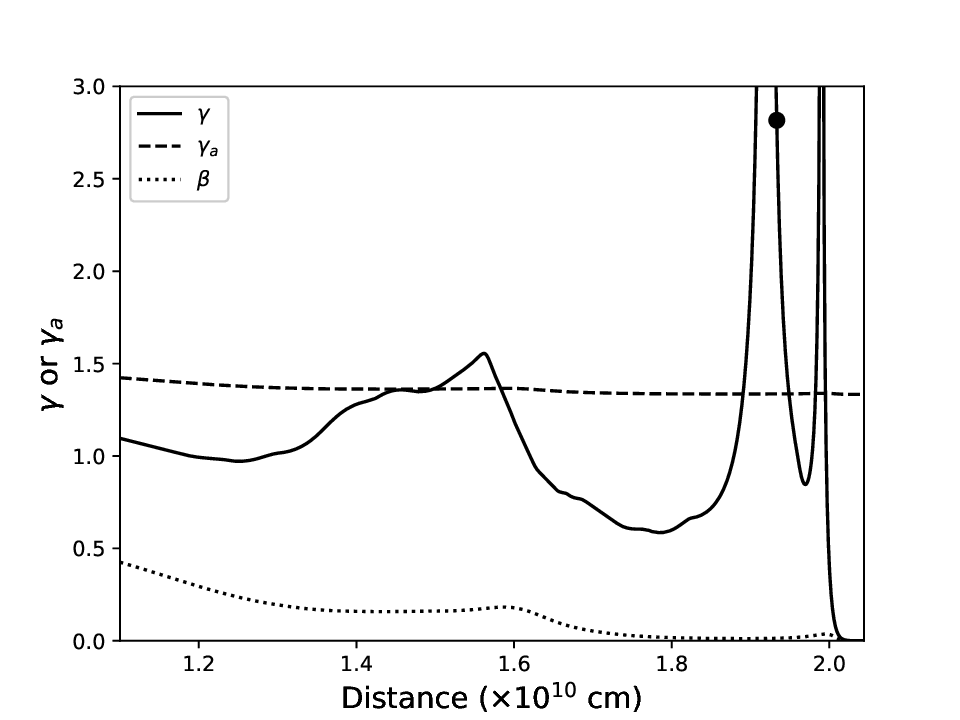}
  \caption{ { Examining  Schwarzschild criterion of the convection in the envelope. The model parameters  are the same as Figure~\ref{fig:ga}, but
    the effect of the spin on the hydrostatic  structure is taken into account. }}
  \label{fig:ga1}
  \end{figure}

 In this study, we have excluded the effects of the spin and convection motion on the stellar structures to focus on
the  basic picture  of the spin evolution under the influence of the MHD wind. To examine   the influence of the stellar spin, we may  solve
the hydrostatic equation of
\begin{equation}
    \frac{dP}{dr}=-\frac{Gm\rho}{r^2}+\rho r\Omega_{WD}^2,
  \end{equation}
where the second term in the right hand side accounts for  the effect of the spin. This form of the hydrostatic equation aligns more consistently with the function $F(r)$ defined by equation~(\ref{eq:func}), which was used to determine the location of the slow-point.

Figure~\ref{fig:ga1} shows the profiles of the index $\gamma=({\rm dlog}P/{\rm dlog}\rho)_{\rm star}$ (solid line) and the adiabatic index $\gamma_a$
(dashed line) near the stellar surface.  The model parameters are identical to those used in  Figure~\ref{fig:ga}. By comparing Figure~\ref{fig:ga1} with
Figure~\ref{fig:ga}, we find that, as expected, the influence of the stellar spin results in an elongation of the stellar structure; the stellar radius is $R_{WD}\sim 2\times 10^{10}$~cm in  Figure~\ref{fig:ga1}, while $\sim 1.6\times 10^{10}$~cm in Figure~\ref{fig:ga}.  Additionally 
the stellar spinning shifts the location of the slow-point (filled circle) outward,  closer to the stellar surface, and results in the decrease of 
the mass-loss rate, as indicated by equation~(\ref{eq:massloss}). As a result,  the influence  of the stellar spin would prolong
the spin-down timescale of the merger product.

Figure~\ref{fig:ga1} demonstrates that the slow-point is located at the convection region, where $\gamma>\gamma_a$. Since the convection alters the temperature profile and hence the profile of the function $F(r)$
in the outer magnetosphere, it will also affect the location of  the slow-point. Consequently, although we would expect that the basic picture of the spin-down of the merger-product is not altered,
a self-consistent treatment of the coupling between the stellar spin and the stellar structure including  convection  will be necessary for subsequent study.

\section{Summary}
\label{sec:5}
The recent observations have collected evidence that the DWD merger event results in the formation of massive and magnetic WD.  In this paper, we have carried out modeling for the spin evolution of merger product that experiences a giant phase,   under the influence of 
MHD wind. We solved the hydrostatic equilibrium of the merger product and determined the slow-point and  Alf\'{v}en-point of the MHD wind.
By assuming that the merger product is born with a luminosity of the order of, but less that  the Eddington value, we estimate a mass-loss rate of $M_{\odot}\sim 10^{20-21}~{\rm g~s^{-1}}$ by the wind. The spin-down timescale is of the order of $\tau_{W}\sim 10$~years when
the magnetic flux of the merger product is $\Phi_B\sim 10^{26}~{\rm G~cm^2}$ or $\tau_{W}\sim 10^3$~years when $\Phi_B\sim 10^{24}~{\rm G~cm^2}$.
We divided the spin-down evolution into three types depending on the magnetic flux  of the merger product. For the strongly magnetized
case,  the initial spin-down timescale is smaller than the cooling timescale of the merger product, and the final angular momentum
is approximately determined by the equation~(\ref{eq:final}) with the initial luminosity.  For mildly magnetized case,
the merger product can be cooled down before the MHD wind is terminated. The final angular momentum is determined when
the spin-down timescale is equal to the cooling timescale. For weaker magnetic case, the initial spin-down timescale is too long and the MHD wind hardly affects to the angular momentum of the merger product. We speculated that the influence of the Carbon shell-burning can be important for the mildly magnetized case,  
for which the spin-down timescale as a result of the MHD wind is of the order of $10^3$~years. We applied our model to three  magnetic WDs, J1901+1458, J2211+1136  and PG~1031+234. We concluded that J1901+1458 and PG~1031+234 corresponds
to the strongly  magnetized case, and their  current spin-periods were almost determined by the initial luminosity. In contrast,
J2211+1136 corresponds to a mildly magnetized case. Since the  current model overlooked  several physical processes,
such as the heat transfer from the envelope to the core, energy transfer by the convection and the shell-burning process, further developing model will be necessary for comprehensive understanding  for the spin evolution of the merger product and the variety of the spin periods of known  magnetized WDs.

\section*{Acknowledgement}
We thank to referee for his/her useful comments and suggestions. We acknowledge discussion with Drs S.Kisaka and  Y.C. Zou and J.T. are supported by the National Key Research and Development Program of China (grant No. 2020YFC2201400) and the National Natural Science Foundation of China (grant No. 12173014).

\section*{Data Availability}
This is a theoretical paper, mainly analytical. All the formulas are available in the article.


\bibliography{sample631}{}
\bibliographystyle{mnras}



\end{document}